\pdfoutput=1
\documentclass[journal]{IEEEtran}
\usepackage{cite}
\usepackage{latexsym}
\usepackage{amssymb}
\usepackage{amsmath}
\usepackage{booktabs}
\usepackage{enumitem}
\usepackage{graphicx}
\usepackage{adjustbox}
\usepackage{xcolor}
\usepackage{url}
\usepackage{algorithm}
\usepackage{algpseudocode}
\usepackage[hidelinks]{hyperref}
\hypersetup{pdftitle={PanoDiff-SR: Synthesizing Dental Panoramic Radiographs using Diffusion and Super-resolution},
  pdfauthor={Sanyam Jain, Bruna Neves de Freitas, Andreas Basse-O'Connor, Alexandros Iosifidis, Ruben Pauwels},
  pdfkeywords={Panoramic Radiography, Diffusion, Super Resolution}}
\graphicspath{{figures/}}
\begin{document}
\title{PanoDiff-SR: Synthesizing Dental Panoramic Radiographs using Diffusion and Super-resolution}

\author{Sanyam~Jain, Bruna~Neves~de~Freitas, Andreas~Basse-O'Connor, Alexandros~Iosifidis, and~Ruben~Pauwels%
\thanks{S. Jain, B. Neves de Freitas and R. Pauwels are with the Department of Dentistry and Oral Health, Aarhus University, 8000 Aarhus, Denmark.}%
\thanks{B. Neves de Freitas is also with the Aarhus Institute of Advanced Studies, Aarhus University, 8000 Aarhus, Denmark.}%
\thanks{A. Basse-O'Connor is with the Department of Mathematics, Aarhus University, 8000 Aarhus, Denmark.}%
\thanks{A. Iosifidis is with the Faculty of Information Technology and Communication Sciences, Tampere University, 33100 Tampere, Finland.}%
\thanks{Corresponding authors: Sanyam Jain (e-mail: sanyam.jain@dent.au.dk) and Ruben Pauwels (e-mail: ruben.pauwels@dent.au.dk).}}

\maketitle

\begin{abstract}
There has been increasing interest in the generation of high-quality, realistic synthetic medical images in recent years. Such synthetic datasets can mitigate the scarcity of public datasets for artificial intelligence research, and can also be used for educational purposes. Work on synthesizing dental panoramic radiographs (PRs) has so far relied almost exclusively on generative adversarial networks (GANs), and three limitations recur: the synthesized field of view is often restricted to a cropped dentoalveolar region rather than a complete PR; observer studies of GAN-generated PRs report that clinicians still identify synthetic images with accuracies of roughly 78--82\%, implying persistent visible giveaways; and adversarial training remains sensitive to instability and mode collapse. Generating a full-size PR directly at diagnostic resolution with a diffusion model is an alternative, but the cost of sampling at that resolution is substantial. In this paper we therefore present PanoDiff-SR, an end-to-end two-stage pipeline developed and evaluated specifically for panoramic radiography: a denoising diffusion implicit model (PanoDiff) generates a low-resolution (LR) seed PR ($256 \times 128$), which a fine-tuned hybrid attention transformer (HAT) super-resolution (SR) stage upscales to a high-resolution (HR) PR of $1024 \times 512$. Decoupling the expensive generative step from the resolution step allows complete, full-arch PRs to be produced at diagnostic resolution on a single 47~GB GPU. Experimental results demonstrate a Fréchet inception distance score of 40.5 between 7243 real and synthetic images (in HR). Inception scores were 2.68, 2.38, 2.90 and 2.98 for real HR, synthetic HR, real LR and synthetic LR images, respectively. Among a diverse group of six clinical experts, all evaluating a mixture of 100 synthetic and 100 real PRs in a time-limited observation, the average accuracy in distinguishing real from synthetic images was 68.5\% (with 50\% corresponding to random guessing), substantially closer to chance than the accuracies reported for GAN-based studies.
\end{abstract}


\section{Introduction}
Panoramic radiography (PR) is a commonly used imaging technique in dentistry that provides an overview of the entire oral cavity, including the teeth, jaw, and surrounding structures \cite{rozylo2021panoramic}. It plays a key role in the diagnosis and treatment planning of various dental conditions such as tooth decay, impacted teeth, and bone abnormalities. As such, there is a strong focus on the automated analysis of PRs using machine learning (ML), e.g. for segmentation\cite{lee2020application} and lesion detection \cite{boztuna2024segmentation}. However, in addition to the challenges related to obtaining high-quality PRs mentioned in \cite{rozylo2021panoramic}, there are few public dental datasets\cite{uribe2024publicly}; furthermore, such datasets tend to be specific for a specific PR device and patient population, and are unlikely to contain rare pathological conditions. Synthetic data can offer a solution to these issues, as they can be pooled with real data to train more robust and generalizable ML models, for example by focusing on underrepresented classes. Furthermore, synthetic images can serve as valuable teaching tools in radiology, allowing for a more personalized approach and ensuring that clinical professionals are familiarized with a wide range of pathologies without relying on patient data. 

Recent studies in generative artificial intelligence (AI) proposed synthesizing high-quality images using denoising diffusion models \cite{ho2020denoising}. Novel diffusion models have achieved groundbreaking results for various applications, such as image inpainting, class-conditional semantic synthesis and super-resolution (SR) \cite{rombach2022high}, license plate image generation \cite{shpir2024license}, panoramic image generation \cite{zhou2024twindiffusion}, underwater image restoration \cite{nathan2024osmosis}, medical image synthesis \cite{muller2023multimodal} and optical geometry control-based generation \cite{voynov2024curved}. However, standard diffusion models are not inherently designed to account for domain-specific constraints, which is crucial for medical imaging applications like the synthesis of radiographs \cite{khalil2025synthetic,mohammad2026generative}. PRs can be particularly challenging in that sense, because of their relatively unique imaging geometry and wide coverage of anatomical and pathological features.  

\begin{figure*}[tp]
    \centering
    \setlength{\tabcolsep}{0pt} %
    \begin{tabular}{cc}
        \includegraphics[width=0.41\textwidth]{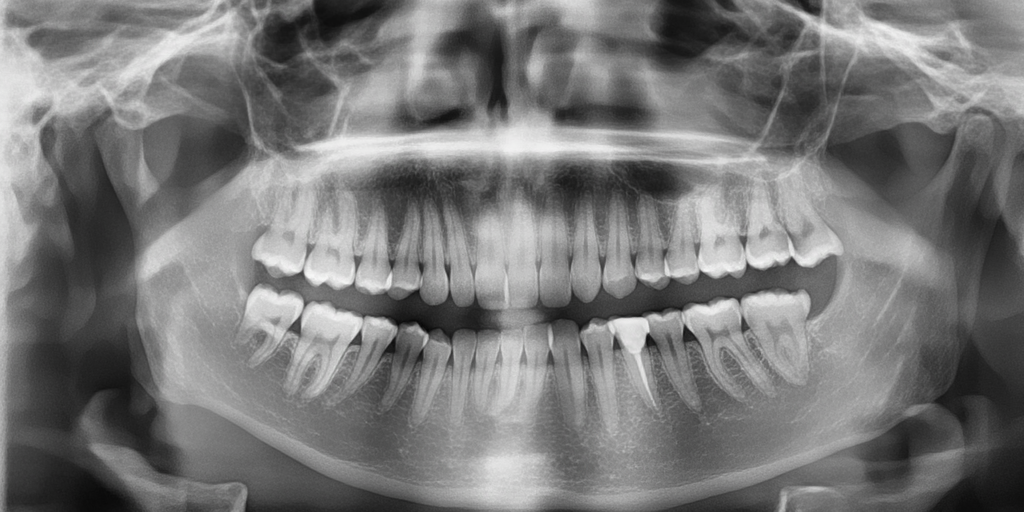} &
        \includegraphics[width=0.41\textwidth]{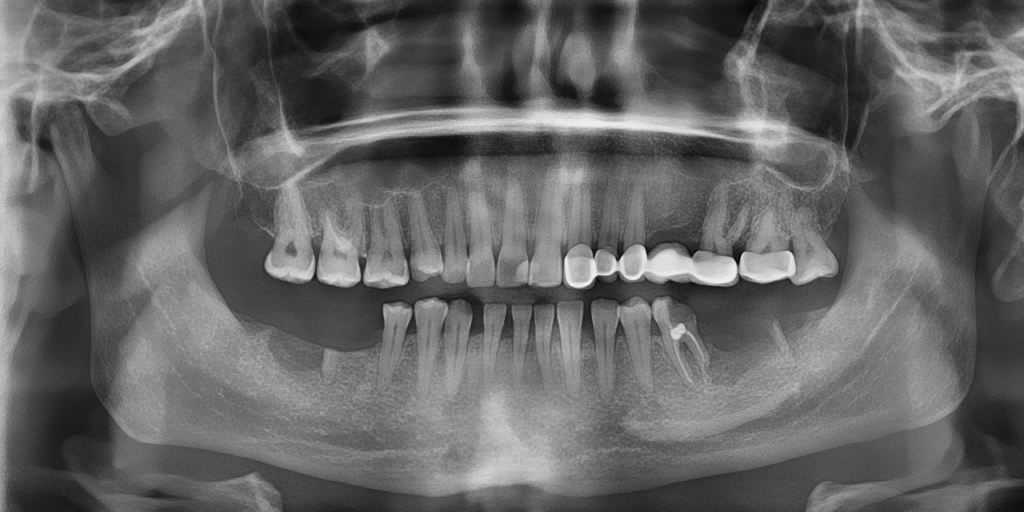} \\
        \includegraphics[width=0.41\textwidth]{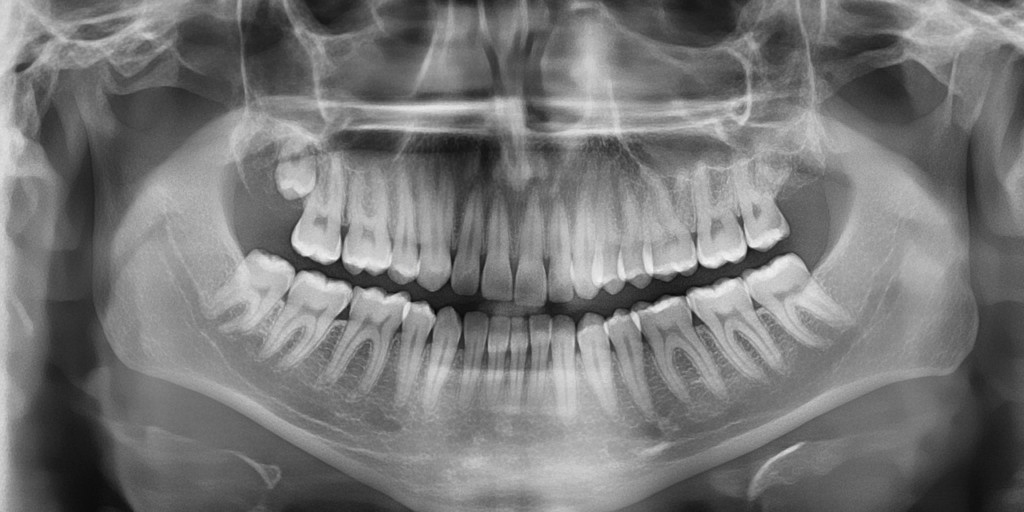} &
        \includegraphics[width=0.41\textwidth]{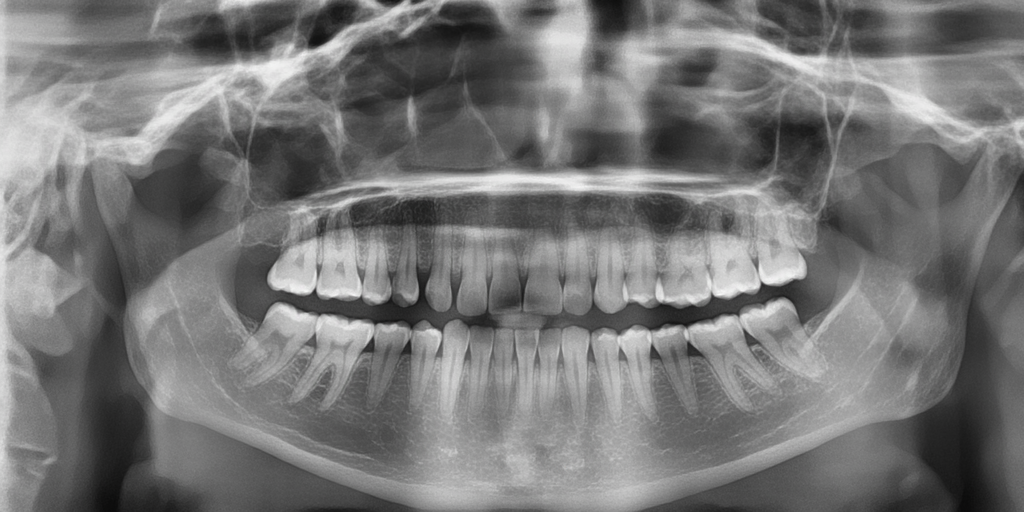} \\
    \end{tabular}
    \caption{Selected synthetic examples from our work that fooled most dentists.}
    \label{fig1realfake}

\end{figure*}

To address this gap, we propose PanoDiff, a diffusion model configured specifically for generating realistic PRs, as shown in Figure \ref{fig1realfake}. Rather than building on generative adversarial networks (GANs) \cite{goodfellow2014generative} or variational autoencoders (VAEs) \cite{kingma2013auto}, for which mode collapse \cite{muller2023multimodal} and blurred outputs \cite{dhariwal2021diffusion} have been reported in medical image synthesis, we employ denoising diffusion implicit models (DDIM) \cite{song2020denoising}, which allow the number of sampling steps to be reduced substantially relative to ancestral sampling of a denoising diffusion probabilistic model (DDPM). This keeps inference tractable, which matters because the model has to be run thousands of times to build a dataset.

After PanoDiff generates a low-resolution (LR) seed PR image of size $256 \times 128$, we employ a SR module to upscale the PRs to $1024 \times 512$, greatly improving the depiction of anatomical details. Unlike traditional upscaling methods that often introduce artifacts or blurriness \cite{wang2018esrgan}, the SR stage uses an existing transformer-based SR network that captures local and global dependencies \cite{chen2023activating,chen2023hat}, which we fine-tune on PR image pairs. Splitting generation from upscaling in this way is what keeps the pipeline affordable: the iterative denoising loop runs only at $256 \times 128$, and the single-pass SR network carries the image to diagnostic resolution.

PanoDiff-SR couples a U-Net denoiser with self-attention, a cosine noise schedule, DDIM sampling and exponential moving average (EMA) weight updates to a hybrid attention transformer (HAT) for super-resolution, with every stage configured for the demands of panoramic radiography. Specifically, this work contributes the following:
\begin{enumerate}
    \item An end-to-end pipeline, PanoDiff-SR, that integrates a DDIM-based diffusion generator with a fine-tuned hybrid attention transformer for super-resolution, together with the domain-specific adaptation and configuration required to make it produce complete, full-arch PRs at $1024 \times 512$ from a pooled multi-device dataset of 7243 images.
    \item A multi-faceted evaluation of the resulting images that combines distribution-level metrics (t-SNE, Fréchet Inception Distance (FID), kernel inception distance (KID), precision and recall, Inception Score (IS)) with a time-limited human observer study \cite{elsayed2018adversarial} involving six dentists of differing seniority, an inter-observer agreement analysis, a Vision Transformer (ViT) trained to separate real from synthetic PRs, and a probe with a medical segmentation foundation model.
    \item A comparison of each stage against alternatives trained on the same data (diffusion versus FastGAN, HAT versus SwinIR), and of the two-stage pipeline against six generators trained directly at full resolution under a comparable budget.
\end{enumerate}

An interactive project page accompanying this work, which presents the generated
radiographs at both resolutions and lets the reader attempt the observer task, is available
at \url{https://s4nyam.github.io/panodiff/}; the implementation is released at
\url{https://github.com/s4nyam/PanoDiff}.

\section{Related Work}
Early research in medical image synthesis primarily relied on GANs, as highlighted in a comprehensive review in \cite{wang2021review}, which summarized advancements up to 2020. This review categorized methods into VAEs, U-Net variants, and GAN-based architectures, noting that GANs dominated due to their ability to generate high-fidelity synthetic images. The study also emphasized clinical applications such as computed tomography (CT) synthesis from magnetic resonance imaging (MRI) and positron emission tomography (PET) image enhancement while acknowledging GAN-related challenges, including training instability and mode collapse. 

Few studies have explored synthesizing PRs using GANs. An exploratory study \cite{pedersen2025pano} trained a GAN with Wasserstein loss and a gradient penalty on cropped PRs, resulting in synthetic PRs with adequate depiction of gross anatomy; however, qualitative analysis indicated a lack of fine details and notable artifacts. Another study attempted to generate smaller portions of PRs with jaw cysts using StyleGAN3 \cite{fukuda2024attempt}; for this specific task, it was shown that oral and maxillofacial radiologists could distinguish real from synthetic images with an accuracy of 82.3\%. Another study used a different version of StyleGAN (StyleGAN2-ADA) to generate full-size PRs \cite{schoenhof2024synthetic}; a large group of medical professionals and students were able to detect synthetic PRs with a sensitivity of 78.2\% and a specificity of 82.5\%. 

Diffusion models have been reported as a more stable alternative to GANs for several medical image synthesis tasks \cite{dayarathna2024deep,kazerouni2023diffusion}, and in the natural-image setting they were shown to surpass strong GAN baselines on established benchmarks \cite{dhariwal2021diffusion}. The Medfusion model \cite{muller2023multimodal} introduced a latent conditional DDPM \cite{ho2020denoising} and reported that, across multiple radiology datasets including CheXpert \cite{irvin2019chexpert}, it achieved lower FID scores and higher precision and recall than the GAN baselines it was compared against, among them StyleGAN-3 \cite{karras2021alias}, ProGAN \cite{karras2017progressive} and a conditional GAN. These results, obtained across several radiology datasets, motivate our choice of a diffusion backbone for PR synthesis. A practical drawback of standard DDPMs is slow inference, since the iterative denoising process requires hundreds to thousands of steps. To address this, we use DDIMs \cite{song2020denoising}, which reduce the number of sampling steps substantially while maintaining image quality. Taken together, these considerations motivate the diffusion-based generator adopted in PanoDiff.

Diffusion-based synthesis of dental radiographs is an active area in its own right, and several approaches complement the GAN-based work described above. Kirkwood et al. \cite{kirkwood2025diffusion} fine-tuned a diffusion model on a deliberately small, expert-curated set of dental radiographs and had the outputs rated by dentists, showing that clinically realistic images can be obtained without a large corpus. Yu et al. \cite{yu2025toothmaker} proposed ToothMaker, which introduces disentangled style and layout controllers so that panoramic radiographs can be generated under explicit anatomical control, and Wang et al. \cite{wang2025difftrgn} applied diffusion with multimodal clinical guidance to tooth root generation. Beyond dentistry, segmentation-guided diffusion has been used to impose anatomical constraints on the generated image \cite{konz2024anatomically}, and broad surveys of diffusion models in medical imaging \cite{kazerouni2023diffusion,dayarathna2024deep} catalogue the same trend across modalities. Relative to these works, the study reported here is unconditional: PanoDiff samples complete PRs without layout, text or segmentation conditioning, which makes it simpler and cheaper to train but gives no control over which anatomy or pathology appears in a given sample. Adding such control is the natural next step.

\section{Material and Methods}
\subsection{Datasets and preparation}
\label{sec:datasets}
\begin{figure*}[tp]
    \centering
    \includegraphics[width=0.5\linewidth]{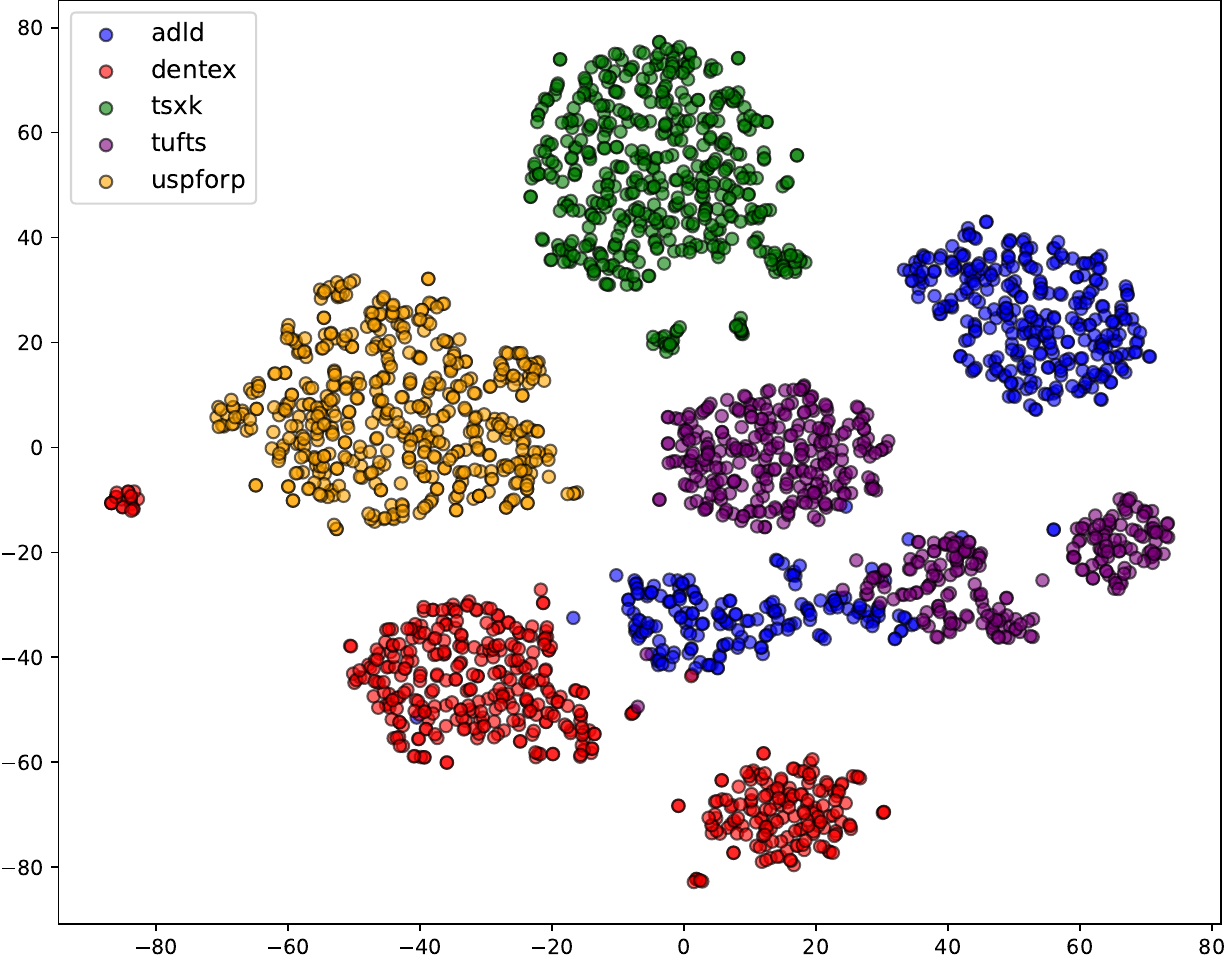}
    \caption{T-distributed stochastic neighbor embedding (t-SNE) plot of 500 random images picked from each source dataset.}
    \label{fig:tsnefolders}
\end{figure*}

In this work, we used five public datasets of PRs as shown in Table \ref{tab:datasetoverview}. The feature space for these datasets, visualized through t-distributed stochastic neighbor embedding (t-SNE), indicates distinct clusters (Figure \ref{fig:tsnefolders}); this can be attributed to slight or moderate variations in image quality (i.e., sharpness, contrast, noise) and anatomical coverage between PR devices from different vendors. All datasets were processed individually by cropping out any overlays or annotations, and resizing to the target resolution ($1024 \times 512$) using the LANCZOS \cite{li2023learning} resampling method. The crop is a fixed margin of 64, 127, 90 and 127 pixels (top, left, bottom and right) removed from every source radiograph before the resize; it is stated here because the same preprocessing is applied to the held-out radiographs of Section~\ref{sec:experimental} and to the real radiographs of Section~\ref{sec:medsam}, which must be framed as the training images are for those comparisons to be meaningful. After the screening described next, the processed images form our training corpus of 7243 files in PNG format.

The image counts in Table~\ref{tab:datasetoverview} are the radiograph files we used from each release. For DENTEX we used the 3603 training images (containing 2686 unique radiographs; the challenge's 50 validation and 250 test images were not used). For TSXK we used all 598 radiographs of the dataset, which entered the corpus twice, as 1196 files. The USPFORP dataset has 72 repeated images among its 945 radiographs. The corpus of 7243 files therefore contains 5653 distinct radiographs. Every quantity in this paper that involves the real set, from the FID and IS values to the classifier of Section~\ref{sec:attention}, was computed on the 7243 files, so a repeated radiograph counts more than once; the distinct counts are given so that the set can be reproduced exactly.

Rejected samples were identified by a single visual pass over all downloaded images, performed by one observer (the first author) at full resolution. An image was excluded if any of the following was clearly visible without ambiguity, i.e. if the decision required no judgement call:
\begin{enumerate}[label=(\roman*)]
    \item severe geometric distortion of the panoramic image, to the extent that the dental arch was no longer continuously traceable;
    \item large machine-generated annotation burnt into the image area, including acquisition dates, patient or clinic identifiers, and left/right (``L''/``R'') orientation markers, where the annotation could not be removed by cropping without also removing dentition;
    \item device-generated artifacts covering a substantial part of the image, such as reconstruction bands, saturated regions or truncation of the image matrix.
\end{enumerate}
These criteria are deliberately conservative: they target images that are unusable as training targets rather than images that are merely of low quality, and a clear distinction was maintained between them and the clinically routine artifacts described next, which were retained. Because each criterion is defined on a plainly visible image property rather than on a judgement of diagnostic quality, the screening can be reapplied directly to these or to further source datasets.

Some of the images in the processed datasets were found to show distinct artifacts, such as (1) metal earrings and ornaments, (2) distortion, and (3) improper patient positioning. We opted to keep these images in the dataset, as such artifacts are considered relatively common and therefore representative of clinical data. In other words, mild to moderate instances of positioning error and distortion were kept, whereas the severe cases described in criterion~(i) above were removed; the consequences of retaining these images for the appearance of the synthetic PRs are discussed in Section~\ref{sec:expert}. For SR training, we downscaled the final processed images fourfold ($256 \times 128$), thus having corresponding low-resolution pairs for each high-resolution (HR) image. These fixed pairs are the real low-resolution set of Section~\ref{sec:lrsr} and the training pairs of the SwinIR comparison; the HAT generator itself was fine-tuned on pairs drawn from the two-stage degradation pipeline of Section~\ref{sec:sr}, which varies the downscaling factor and adds blur, noise and compression.

\begin{table*}[tp]
    \centering
    \caption{Overview of dental radiography datasets used in our study. Dataset abbreviations with format, year, and availability: ADLD (png, 2024, Kaggle)*, DENTEX (png, 2023, Zenodo), TSXK (png, 2023, Kaggle), TUFTS (jpg, 2022, On Request), USPFORP (jpg, 2024, On Request). * indicates that the dataset was recently updated with 1500 more images, but we accessed it when it had 500 images. The 'Images' column denotes the radiograph files used from each release. $\sim$ indicates varying sizes in the dataset within the given resolution range. Abbreviations: ADLD – A dual-labeled dataset, DENTEX – Dental Enumeration and Diagnosis on Panoramic X-rays, TSXK – Teeth Segmentation on dental X-ray images, TUFTS – Tufts Dental Database, USPFORP – São Paulo dataset.}
    \label{tab:datasetoverview}
   
    \begin{tabular}{|c|c|c|c|}
        \hline
        \textbf{Abbr.} & \textbf{Images} & \textbf{Country} & \textbf{Resolution} \\ 
        \hline
        ADLD \cite{zhou2024dual} & 500* & China & $\sim$2940$\times$1435 or $\sim$987$\times$478 \\ 
        \hline
        DENTEX \cite{hamamci2023dentex} & 3603 & Switzerland & $\sim$2950$\times$1316 or $\sim$1976$\times$976 \\ 
        \hline
        TSXK \cite{humans_in_the_loop_2023} & 598 ($\times 2$) & DR Congo & 2041$\times$1024 \\ 
        \hline
        TUFTS \cite{panetta2021tufts} & 1000 & USA & 1615$\times$840 \\ 
        \hline
        USPFORP \cite{costa2024development} & 945 & Brazil & 2903$\times$1536 \\ 
        \hline
    \end{tabular}
\end{table*}

\subsection{PanoDiff}
\begin{figure*}[tp]
    \centering
    \includegraphics[width=0.5\linewidth]{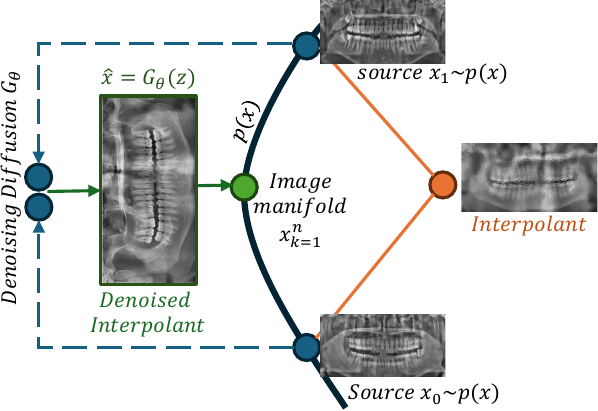}
    \caption{General principle of synthetic image generation through manifold representation. Consider a dataset of images \( \{x_k\}_{k=1}^n \), where \( x_k \sim p(x) \). These images serve as samples from the target distribution \( p(x) \). A best sampler \( G_\theta \) is one such that $\hat x$ \( = G_\theta(z) \), where \( z \sim \mathcal{N}(0, I) \), to produce high-quality samples resembling the true data distribution \( p\). }
    \label{fig:manifold}
\end{figure*}

PanoDiff is a diffusion-based generative model designed to synthesize low-resolution PRs of size $256 \times 128$. The top-level principle of PanoDiff, illustrated in Figure \ref{fig:manifold}, involves generating synthetic images by sampling from a manifold representation using a trained sampler $G_\theta$ that transforms noise $z \sim N(0, I)$ into high-quality images resembling the target distribution $p(x)$.

Built on DDPMs, PanoDiff employs a U-Net architecture enhanced with self-attention mechanisms and residual blocks \cite{vaswani2017attention}. The forward process gradually adds Gaussian noise to the original image $x_0$ over $T$ time steps, producing a noisy image $x_t$ as described in Equation~\ref{eq:forward_process}:
\begin{eqnarray}\label{eq:forward_process}
x_t & = & \sqrt{\alpha_t} x_0 + \sqrt{1 - \alpha_t} \epsilon_t, \quad \epsilon_t \sim \mathcal{N}(0, I),
\end{eqnarray}
where $\alpha_t$ is the cumulative product of noise scheduling coefficients. The reverse process, guided by a learned noise prediction network $\epsilon_\theta(x_t, t)$, iteratively denoises $x_t$ to recover $x_0$. This approach ensures high-fidelity image generation while avoiding common pitfalls such as mode collapse and blurry outputs.

The U-Net in PanoDiff consists of hidden layers with dimensions 64, 128, 256, 512, which progressively downsample and upsample the input image while maintaining feature richness at each resolution level. This hierarchical structure is particularly effective for handling the complexity of PRs, as it ensures that both low-level textures and high-level structures are preserved during the denoising process. The self-attention mechanism, integrated into the U-Net is particularly useful in PanoDiff, as it allows the model to capture long-range dependencies across the PR, ensuring that anatomical structures remain spatially coherent during generation.

Because the width specification alone leaves the spatial behaviour of the network ambiguous, we document the encoder and decoder in full. The network operates on tensors of shape $3 \times 128 \times 256$ ($C \times H \times W$; the grayscale PR is replicated across three channels; spatial sizes in this subsection are therefore height $\times$ width, whereas images elsewhere in the paper are given as width $\times$ height, e.g.\ $256 \times 128$) and returns a noise estimate of the same shape. It has $3.40 \times 10^{7}$ trainable parameters. The configuration is as follows.

\emph{Stem and conditioning.} A $3 \times 3$ convolution with stride~1 and padding~1 maps $3 \rightarrow 64$ channels at full $128 \times 256$ resolution. The timestep $t$ is encoded by a sinusoidal embedding of dimension 64, followed by $\mathrm{Linear}(64 \rightarrow 256)$, SiLU and $\mathrm{Linear}(256 \rightarrow 256)$. Every residual block receives this 256-dimensional embedding through a SiLU and a linear projection to its own channel count, added as a per-channel bias between the block's two convolutions.

\emph{Residual block.} Each residual block applies $3 \times 3$ convolution (stride~1, padding~1), $\mathrm{GroupNorm}$ with 8 groups and SiLU; then adds the timestep projection; then repeats convolution, normalisation and activation. The identity path is a $1 \times 1$ convolution when the channel count changes and an identity mapping otherwise.

\emph{Encoder.} Three stages are applied, each consisting of two residual blocks at constant width, an optional self-attention layer, and a transition:
\begin{enumerate}[label=(\arabic*)]
    \item 64 channels at $128 \times 256$, no attention; transition to 128 channels at $64 \times 128$;
    \item 128 channels at $64 \times 128$, with attention; transition to 256 channels at $32 \times 64$;
    \item 256 channels at $32 \times 64$, with attention; transition to 512 channels, \emph{without} any change of spatial size.
\end{enumerate}
The transitions in stages~(1) and~(2) are a pixel-unshuffle (space-to-depth) rearrangement with factor 2 followed by a $1 \times 1$ convolution that maps the $4\times$ inflated channel count to the target width. The transition in stage~(3) is a plain $3 \times 3$ convolution with stride~1, i.e. a channel projection only. Strided convolutions and pooling are not used anywhere in the network.

\emph{Bottleneck.} A residual block, a self-attention layer and a second residual block, all at 512 channels and $32 \times 64$.

\emph{Decoder.} Three stages mirror the encoder. Each stage concatenates the two skip tensors saved by the corresponding encoder stage before its two residual blocks, applies attention in the first two stages, and then upsamples by nearest-neighbour interpolation with factor 2 followed by a $3 \times 3$ convolution; the final decoder stage uses a $3 \times 3$ convolution without spatial change, mirroring encoder stage~(3). Output is produced by a residual block on the concatenation of the decoder output with the stem features ($128 \rightarrow 64$ channels) and a final $1 \times 1$ convolution to 3 channels.

\emph{Attention.} All self-attention layers use 4 heads with 32 channels per head, computed by $1 \times 1$ convolutions for the queries, keys and values, with RMS normalisation applied before the projection and a residual connection around the layer.

Two consequences of this design are worth stating explicitly, as they bear on how much spatial detail survives the encoder. First, the network performs only \emph{two} halvings of the spatial resolution, so the coarsest feature map is $32 \times 64$, which is one quarter of the input resolution along each axis and not the $8 \times 16$ that four successive halvings would give. Second, those two reductions are pixel-unshuffle rearrangements rather than strided convolutions, so the reduction step itself moves information from the spatial axes into the channel axis instead of discarding it. Attention is deliberately omitted at the finest $128 \times 256$ level, where a dense attention map would dominate memory, and applied at all coarser levels where the long-range dependencies along the dental arch are represented.

\begin{figure*}[tp]
    \centering
    \includegraphics[width=\linewidth]{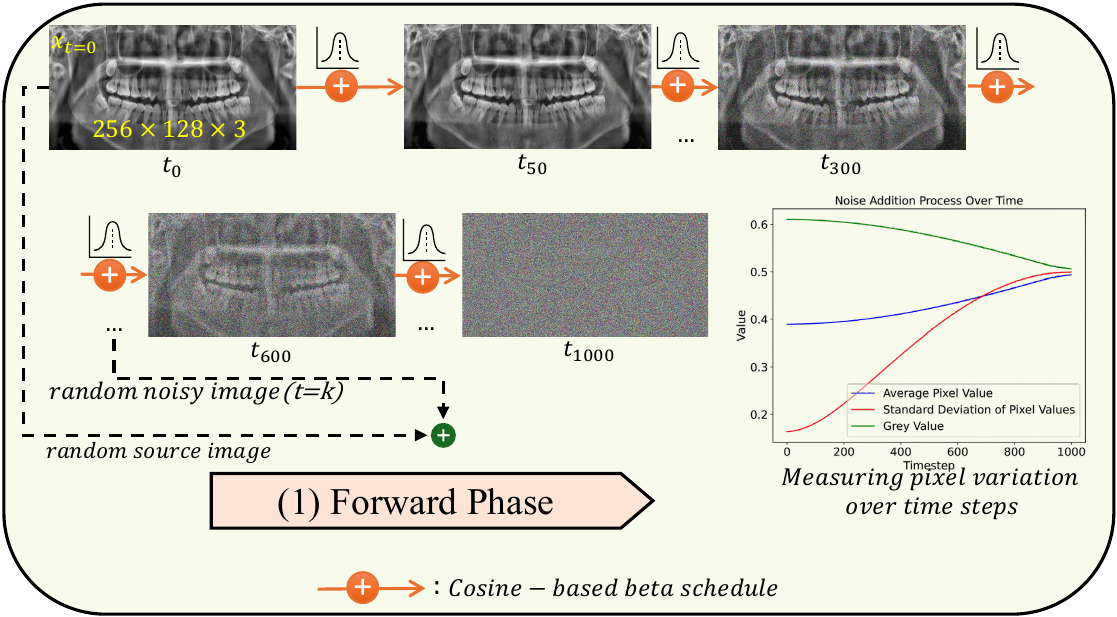}
    \caption{Forward phase of PanoDiff. Noise is added to the input image $x_{t=0}$ over $T = 1000$ time steps following the cosine $\beta$-schedule, until the image is pure noise at $t = 1000$. The plot at the lower right tracks the pixel statistics along the trajectory.}
    \label{fig:panodiff-forward}
\end{figure*}

PanoDiff uses a cosine-based beta schedule for noise addition, defined in Equation~\ref{eq:beta_schedule}:
\begin{eqnarray}\label{eq:beta_schedule}
\beta_t & = & \text{clip}\left(1 - \frac{\alpha_t}{\alpha_{t-1}}, \beta_{\text{min}}, \beta_{\text{max}}\right),
\end{eqnarray}
where $\alpha_t$ is derived from a cosine function\footnote{In the forward process, the cosine schedule adds noise more gradually at the start and more aggressively later. For the reverse process, it decreases noise more gradually in the beginning and more aggressively towards the end using a cosine function.} as described in Figure~\ref{fig:panodiff-forward}.

\begin{figure*}[tp]
    \centering
    \includegraphics[width=\linewidth]{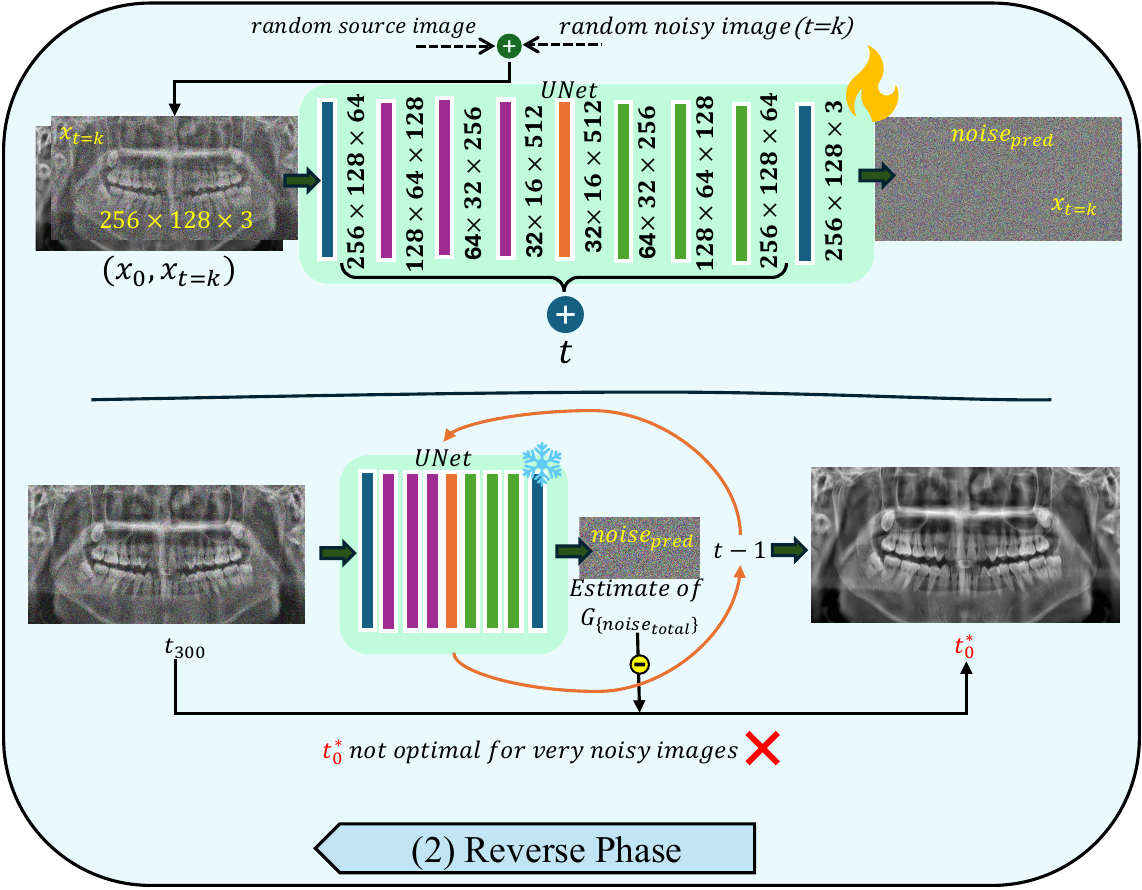}
    \caption{Reverse phase of PanoDiff. Top: the U-Net (flame: trainable) is trained on $(x_0, x_{t=k})$ with an $L_1$ loss to predict the noise contained in $x_{t=k}$. Bottom, for comparison: step-by-step DDPM denoising with the frozen network (snowflake), for which a single-step estimate is not adequate at high noise levels.}
    \label{fig:panodiff-reverse}
\end{figure*}

\begin{figure*}[tp]
    \centering
    \includegraphics[width=\linewidth]{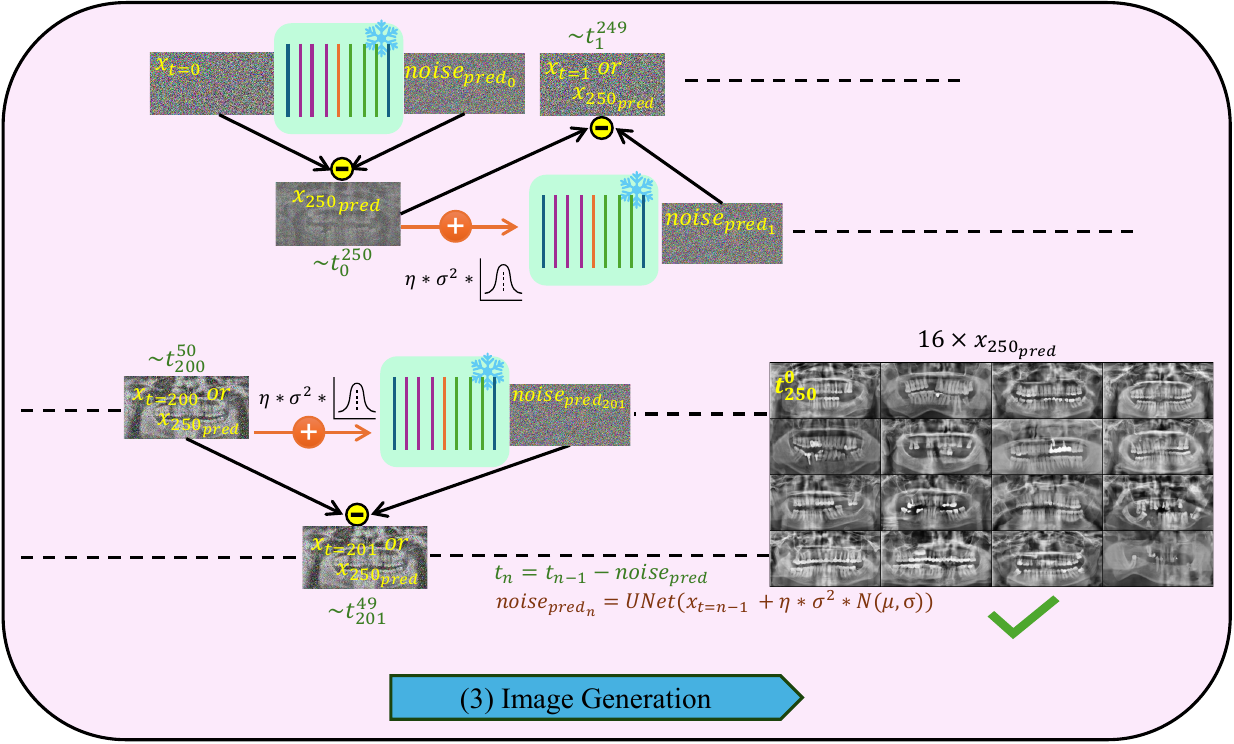}
    \caption{Image generation in PanoDiff. Starting from noise, the frozen U-Net predicts and removes noise, noise is added back, and the loop repeats for the number of inference time steps. Sixteen generated PRs are shown at the lower right.}
    \label{fig:panodiff-generation}
\end{figure*}

The three phases are shown separately in Figures~\ref{fig:panodiff-forward},
\ref{fig:panodiff-reverse} and~\ref{fig:panodiff-generation}, one figure per phase, and
the explanation that accompanies them is given here rather than in the captions.

\emph{Forward phase} (Figure~\ref{fig:panodiff-forward}). Gaussian noise is added to the
input image $x_{t=0}$ over $T = 1000$ time steps under the cosine $\beta$-schedule of
Equation~\ref{eq:beta_schedule}, which adds noise gradually at the start and more
aggressively towards the end. By $t = 1000$ the image is indistinguishable from noise,
which is what the pixel statistics plotted at the lower right of the figure show: the mean
pixel value, the standard deviation of the pixel values and the grey value converge as
the signal disappears. Training does not traverse the whole trajectory. It draws a random
source image together with a random noisy image at a single timestep $t = k$, which is
the path drawn as a dashed line in the figure.

\emph{Reverse phase} (Figure~\ref{fig:panodiff-reverse}). The U-Net is trained on the pair
$(x_0, x_{t=k})$, conditioned on the timestep $t$, with an $L_1$ loss, to predict the
noise that $x_{t=k}$ contains; the tensor shapes documented above are annotated along the
encoder and decoder in the figure, and the flame marks the network as trainable. The
lower part of the figure contrasts this with ancestral DDPM sampling, in which the
frozen network (snowflake) removes noise one step at a time. A single-step estimate
$t_0^{*}$ is not adequate for a very noisy image, which is why generation has to iterate.

\emph{Image generation} (Figure~\ref{fig:panodiff-generation}). Sampling starts from pure
noise. The frozen U-Net predicts the noise, which is subtracted to give a slightly less
noisy image; noise scaled by $\eta \sigma^2$ is then added back and the result is fed to
the network again. The loop repeats for the number of inference time steps, which is 250
under the DDIM scheduler introduced below rather than the 1000 steps of the forward
process. Sixteen PRs produced by this procedure are shown at the lower right of the
figure.

This schedule, introduced in \cite{nichol2021improved}, ensures smoother transitions between noise levels. Furthermore, the model employs a DDIM scheduler during inference, which reduces the number of required denoising steps from $T=1000$ to $T=250$. The DDIM update rule is given by Equation~\ref{eq:ddim_update}:
\begin{eqnarray}\label{eq:ddim_update}
x_{t-1} & = & \sqrt{\alpha_{t-1}} \left( \frac{x_t - \sqrt{1 - \alpha_t} \epsilon_\theta(x_t, t)}{\sqrt{\alpha_t}} \right) \notag \\
& & + \sqrt{1 - \alpha_{t-1} - \sigma_t^2} \epsilon_\theta(x_t, t) + \sigma_t z,
\end{eqnarray}
where $z \sim \mathcal{N}(0, I)$ and $\sigma_t$ controls the stochasticity of the process. In our setting this acceleration did not produce a visible loss of image quality, which makes repeated sampling practical. PanoDiff further uses an EMA with a cosine-based update schedule for model weights, a stabilisation technique that is standard in diffusion training rather than specific to this work. The EMA coefficient $\gamma$ is updated as described in Equation~\ref{eq:ema_coefficient}:
\begin{eqnarray}\label{eq:ema_coefficient}
\gamma_k & = & 1 - (1 - \gamma_0) \frac{\cos(\pi k / K) + 1}{2},
\end{eqnarray}
where $k$ is the current training step and $K$ is the total number of steps. This approach stabilizes training by gradually increasing the EMA coefficient, ensuring that the model converges to a robust solution. The AdamW optimizer, combined with gradient clipping and flash attention, further enhances training efficiency and stability.

In summary, PanoDiff combines a U-Net denoiser with self-attention \cite{ho2020denoising,vaswani2017attention}, a cosine beta schedule \cite{nichol2021improved}, DDIM sampling \cite{song2020denoising} and EMA weight averaging, each selected and configured for the specific characteristics of panoramic radiographs: a wide, strongly anisotropic field of view, a single grayscale channel, and a globally consistent arch geometry that has to be preserved across the full width of the image. It is these characteristics, rather than the generic image-synthesis setting, that determine the configuration reported here. The overall functioning of PanoDiff is shown in Figures~\ref{fig:panodiff-forward}--\ref{fig:panodiff-generation}.

\subsection{Super-resolution}
\label{sec:sr}
The SR module in our framework enhances low-resolution PRs to high-resolution images of size $1024 \times 512$. Rather than a CNN-based SR method, we adopt an existing Transformer-based architecture with hybrid attention, HAT, of Chen et al. \cite{chen2023activating,chen2023hat}, without architectural modification. As introduced by its authors, HAT integrates window-based multi-head self-attention with overlapping cross-attention to capture both local and global dependencies, and stacks residual hybrid attention groups (RHAG) to refine features across layers. Our use of it is a domain adaptation: we fine-tune the released weights on PR image pairs so that the network's priors, learned on natural images, are re-tuned to the texture statistics of radiographs.

To simulate real-world degradations, we adopt the two-stage degradation pipeline used for training real-world SR models \cite{chen2023hat}, configured for our data. The first degradation stage applies Poisson noise to the ground-truth images, followed by JPEG compression. The second degradation stage further introduces additional blurring and Gaussian noise. This degradation pipeline is crucial for training the SR model to handle various image quality manifestations from PRs effectively.

To ensure both structural accuracy and perceptual quality, the SR module is trained using a combination of three key losses: pixel-wise loss (\( \mathcal{L}_1 \)), perceptual loss (\( \mathcal{L}_{\text{percep}} \)), and adversarial loss (\( \mathcal{L}_{\text{GAN}} \)). The pixel-wise loss, defined in Equation~\ref{eq:pixel_loss}, ensures pixel-level accuracy by minimizing the \( L_1 \) distance between the synthetic and ground-truth PRs:
\begin{eqnarray}\label{eq:pixel_loss}
\mathcal{L}_1 & = & \| G(x) - y \|_1,
\end{eqnarray}
where \( G(x) \) is the generated high-resolution image and \( y \) is the ground truth. The perceptual loss, computed using a pre-trained VGG network (on ImageNet \cite{deng2009imagenet}) and given by Equation~\ref{eq:perceptual_loss}, ensures that the synthetic PRs are visually realistic by minimizing the difference in feature representations:
\begin{eqnarray}\label{eq:perceptual_loss}
\mathcal{L}_{\text{percep}} & = & \| \phi(G(x)) - \phi(y) \|_2^2,
\end{eqnarray}
where \( \phi \) represents the feature extraction function of the VGG network. Finally, the adversarial loss, implemented using a GAN and defined in Equation~\ref{eq:adversarial_loss}, encourages the generator to produce PRs that are indistinguishable from real high-resolution PRs:
\begin{eqnarray}\label{eq:adversarial_loss}
\mathcal{L}_{\text{GAN}} & = & \mathbb{E}_{x \sim p_{\text{data}}(x)}[\log D(x)] \notag \\
& & + \mathbb{E}_{z \sim p_z(z)}[\log(1 - D(G(z)))],
\end{eqnarray}
where \( G \) is the generator, \( D \) is the discriminator, \( x \) represents real high-resolution PRs, and \( z \) represents low-resolution inputs. The discriminator \( D \) is implemented as a U-Net with spectral normalization (SN). The U-Net architecture of the discriminator consists of an encoder-decoder structure with skip connections, enabling it to capture multi-scale features effectively. SN is applied to stabilize the training process by constraining the Lipschitz constant of the discriminator. This design ensures that the discriminator provides strong feedback to the generator, improving the overall quality of the synthetic PRs.

To further enhance the diversity of training data, we employ a training pair pool in which the downscaling factor is varied between 2, 3 and 4. This pool ensures that the model is exposed to a wide variety of degradation patterns, improving its generalization ability. The total loss for the generator, defined in Equation~\ref{eq:total_loss}, is a weighted combination of the three losses:
\begin{eqnarray}\label{eq:total_loss}
\mathcal{L}_{\text{total}} & = & \lambda_1 \mathcal{L}_1 + \lambda_2 \mathcal{L}_{\text{percep}} + \lambda_3 \mathcal{L}_{\text{GAN}},
\end{eqnarray}
where \( \lambda_1 \), \( \lambda_2 \), and \( \lambda_3 \) are weighting factors. This combination of transformer-based attention, adversarial training and a multi-term loss is the training recipe proposed with HAT, which was reported to outperform earlier SR methods on standard benchmarks \cite{chen2023activating}; we reuse it unchanged apart from the fine-tuning data.

Algorithm~\ref{alg:panodiff-sr} states the complete pipeline in the notation used above:
the two training stages, which are run once, and the synthesis path that produces one
high-resolution PR.

\begin{algorithm*}[tp]
\caption{PanoDiff-SR: training and synthesis}
\label{alg:panodiff-sr}
\small
\begin{algorithmic}[1]
\Require real PR corpus $\mathcal{D}$ at $1024 \times 512$; diffusion steps $T = 1000$;
         sampling steps $S = 250$; stochasticity $\eta = 1$; cosine schedule
         $\{\alpha_t\}_{t=1}^{T}$ (Eq.~\ref{eq:beta_schedule}); upscaling factor $4$
\Ensure  synthetic PR at $1024 \times 512$

\Statex \textbf{Stage 1: train the denoising network $\epsilon_\theta$}
\State initialise $\theta$ at random; $\theta_{\text{EMA}} \gets \theta$
\For{$110$ epochs, minibatches of $4$}
  \State $x_0 \gets$ image from $\mathcal{D}$, resized to $256 \times 128$ and scaled to $[-1,1]$
  \State $t \sim \mathcal{U}\{1,\dots,T\}$, \quad $\epsilon \sim \mathcal{N}(0, I)$
  \State $x_t \gets \sqrt{\alpha_t}\, x_0 + \sqrt{1 - \alpha_t}\, \epsilon$
         \Comment{forward phase, Fig.~\ref{fig:panodiff-forward}}
  \State $\theta \gets \textsc{AdamW}\big(\theta, \nabla_\theta \| \epsilon_\theta(x_t, t) - \epsilon \|_1 \big)$
         \Comment{reverse phase, Fig.~\ref{fig:panodiff-reverse}}
  \State $\theta_{\text{EMA}} \gets \gamma\, \theta_{\text{EMA}} + (1 - \gamma)\, \theta$;
         update $\gamma$ by Eq.~\ref{eq:ema_coefficient}
\EndFor

\Statex \textbf{Stage 2: fine-tune the super-resolution generator $G$}
\State initialise $G$ from the released HAT $4\times$ weights; initialise discriminator $D$
\For{$400{,}000$ iterations, minibatches of $4$}
  \State $y \gets$ crop from $\mathcal{D}$; \; $x \gets \textsc{Degrade}(y)$
         \Comment{two-stage degradation}
  \State update $G$ on $\mathcal{L}_{\text{total}} = \lambda_1 \mathcal{L}_1 + \lambda_2 \mathcal{L}_{\text{percep}} + \lambda_3 \mathcal{L}_{\text{GAN}}$ (Eq.~\ref{eq:total_loss}); update $D$
\EndFor

\Statex \textbf{Stage 3: synthesise one image}
\State $x \sim \mathcal{N}(0, I)$ of size $256 \times 128 \times 3$
\For{$t = T - 4,\, T - 8,\, \dots,\, 0$}
       \Comment{$S$ steps spaced $T/S = 4$ apart}
  \State $\hat{\epsilon} \gets \epsilon_{\theta_{\text{EMA}}}(x, t)$
  \State $\hat{x}_0 \gets \text{clip}\!\left( \dfrac{x - \sqrt{1 - \alpha_t}\, \hat{\epsilon}}{\sqrt{\alpha_t}},\, -1,\, 1 \right)$
  \State $x \gets \sqrt{\alpha_{t-1}}\, \hat{x}_0 + \sqrt{1 - \alpha_{t-1} - \sigma_t^2}\, \hat{\epsilon} + \sigma_t z$,
         \; $z \sim \mathcal{N}(0, I)$
         \Comment{Eq.~\ref{eq:ddim_update}}
\EndFor
\State $I_{\text{LR}} \gets$ rescale $x$ to $[0,1]$
       \Comment{$256 \times 128$, Fig.~\ref{fig:panodiff-generation}}
\State \Return $G(I_{\text{LR}})$ \Comment{$1024 \times 512$}
\end{algorithmic}
\end{algorithm*}

\subsection{Experimental Setup, Implementation and Training}
\label{sec:experimental}
Using 7243 training images of PRs ($H=512, W=1024$), PanoDiff was trained for $110$ epochs. The forward phase was performed for $1000$ steps and reverse phase for $250$ steps. 

For SR we did not modify the HAT architecture or introduce new loss terms. We initialised the generator from the pre-trained HAT model for $4\times$ super-resolution released by Chen et al. \cite{chen2023hat,chen2023activating} and fine-tuned it on our dataset's image pairs, retaining the remainder of the reference implementation. The model was trained on a dataset of high-resolution images with a scale factor of 4, utilizing a combination of L1, perceptual, and GAN losses. The training process involves two degradation stages with various augmentations, including resizing, noise addition, and blurring. The generator (HAT) and discriminator (UNetDiscriminatorSN) are optimized using Adam with a learning rate of $1e-4$, with the training running for 400,000 iterations. Both stages of PanoDiff-SR were trained on a single NVIDIA RTX 6000 Ada Generation GPU with 47~GB of memory, at a batch size of 4, and the synthetic dataset was subsequently generated in batches on the same device; the comparisons of Section~\ref{sec:directhr} were run on a different GPU (Table~\ref{tab:directhr}). The training figures quoted below therefore refer to the RTX 6000 Ada; the inference times were measured for both stages back to back on the GPU of Section~\ref{sec:directhr}, so the two stages can be compared with each other directly. The resulting computational cost is summarised in Table~\ref{tab:efficiency}. Training and
sampling code for both stages, together with the pre-trained checkpoints, are available at
\url{https://github.com/s4nyam/PanoDiff}.

A practical motivation for splitting generation from upscaling is that it keeps the cost of building a synthetic dataset manageable, so we quantify that cost rather than describing it qualitatively. Table~\ref{tab:efficiency} reports, for each stage, the trainable parameter count, the training schedule and its measured wall-clock time, the number of network evaluations required per generated image, and the inference time per image.

Three observations follow. First, the complete pipeline is trained in approximately $70.8$ hours (just under three days) on a single GPU, so reproducing it requires neither multi-GPU nor distributed training. Second, the cost is dominated by the super-resolution stage ($53.1$ h against $17.7$ h), which follows from fine-tuning for 400,000 iterations rather than from the diffusion model being expensive to train. Third, and most relevant to the two-stage design, generation is dominated by the iterative sampler: the 250-step DDIM loop costs about 22 times the single super-resolution pass on the same device, so upscaling adds roughly $4\%$ to the per-image cost while increasing the pixel count sixteenfold. Generating an image directly at $1024 \times 512$ with the same sampler would instead require every one of the 250 denoising steps to operate on sixteen times as many pixels; the two-stage arrangement confines that iterative cost to the low-resolution grid.

At the measured rates, producing the complete set of $7243$ synthetic PRs used throughout this study amounts to roughly $14.4$~h on a single GPU: $13.8$~h of continuous batched sampling and a further $0.6$~h for the super-resolution pass. This is a one-off cost: once generated, the same set is reused for every experiment reported below.

The main design choices were checked against their alternatives. Fine-tuning the SR stage on PR pairs is the most consequential of them: on 185 held-out real PRs (DENTEX validation and test images that are not in the training corpus) downsampled fourfold and restored, it lowers the learned perceptual image patch similarity (LPIPS) \cite{zhang2018unreasonable} from $0.310$ with the released HAT weights to $0.137$. For the diffusion stage, weight averaging is decisive (FID on 1000 samples of $67.8$, against $122.5$ with the raw weights), and the 250-step DDIM trajectory is conservative, since 100 steps sample $2.5$ times faster at a cost of about $5$ FID points.

\begin{table*}[tp]
\centering
\caption{Computational cost of the two stages of PanoDiff-SR. The training figures refer to a single NVIDIA RTX 6000 Ada Generation GPU (47~GB); the inference times were measured for both stages under identical conditions on the GPU used in Section~\ref{sec:directhr}. LR: $256 \times 128$; HR: $1024 \times 512$. The SR stage is fine-tuned on random $256 \times 256$ crops of the real HR radiographs, the standard practice for HAT; since it accepts any image size, it is applied to the whole $256 \times 128$ output of PanoDiff at inference, without padding or rescaling. Each schedule is given in the unit of its training code, with the equivalent in the other unit in brackets.}
\label{tab:efficiency}
\small
\setlength{\tabcolsep}{5pt}
\begin{tabular}{@{}lcc@{}}
\toprule
 & \textbf{PanoDiff} & \textbf{SR (HAT, $4\times$)} \\
 & \textbf{(LR synthesis)} & \textbf{(LR $\rightarrow$ HR)} \\
\midrule
Trainable parameters                       & $34.0$ M          & $20.8$ M \\
Training input                             & $256 \times 128$  & $256 \times 256$ crops \\
Batch size                                 & 4                 & 4 \\
Training schedule                          & 110 epochs        & 400{,}000 iter. \\
                                           & (199{,}210 steps) & ($\approx$221 epochs) \\
Mean time per step                         & $0.32$ s          & $0.48$ s \\
Training wall-clock                        & $17.7$ h          & $53.1$ h \\
Training hardware                          & \multicolumn{2}{c}{one RTX 6000 Ada (47 GB)} \\
\midrule
Network evaluations per image              & 250 (DDIM)        & 1 \\
Inference time per image                   & $6.84$ s          & $0.31$ s \\
\bottomrule
\end{tabular}

\end{table*}
We can clearly observe the progression in anatomical consistency across epochs in Figure \ref{fig:epoch_comparison}. By fixing the random seed, we generate comparable samples at each epoch, allowing us to track the model's refinement over time. Initially, the model captures coarse global structures (jaw shape and mandibular canal) with some degree of blurring, while later epochs produce increasingly anatomically consistent PRs and finer details.

\begin{figure*}[tp]
    \centering
    \setlength{\tabcolsep}{1pt}
    \renewcommand{\arraystretch}{0.7}
    \begin{tabular}{cccccc}
        \textbf{Epoch 11} & \textbf{Epoch 33} & \textbf{Epoch 55} & \textbf{Epoch 77} & \textbf{Epoch 99} & \textbf{Epoch 110} \\
        
        \includegraphics[width=0.16\linewidth]{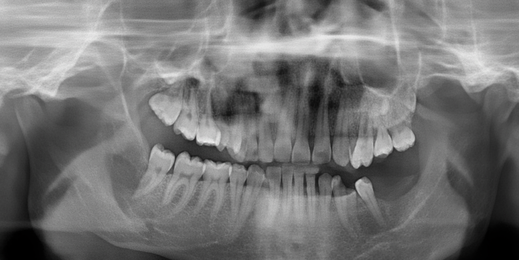} &
        \includegraphics[width=0.16\linewidth]{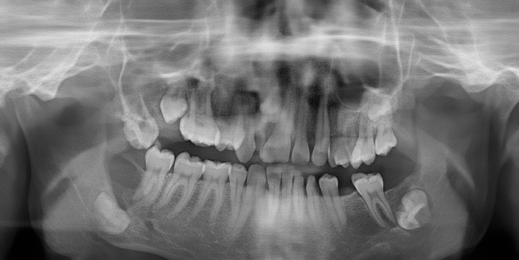} &
        \includegraphics[width=0.16\linewidth]{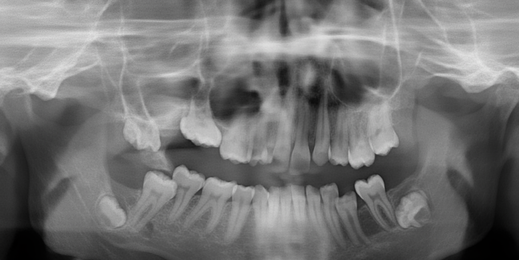} &
        \includegraphics[width=0.16\linewidth]{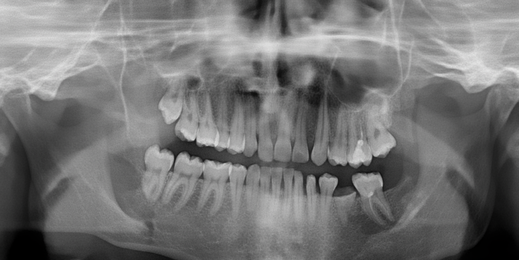} &
        \includegraphics[width=0.16\linewidth]{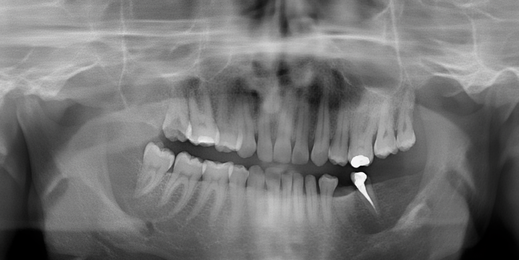} &
        \includegraphics[width=0.16\linewidth]{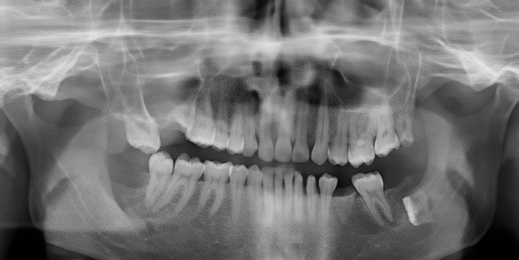} \\
        
        \includegraphics[width=0.16\linewidth]{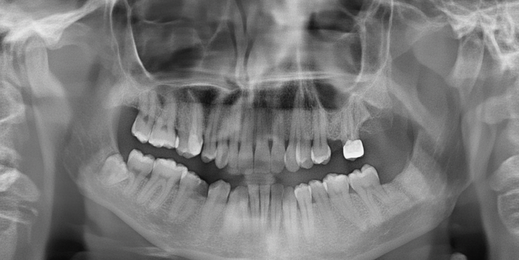} &
        \includegraphics[width=0.16\linewidth]{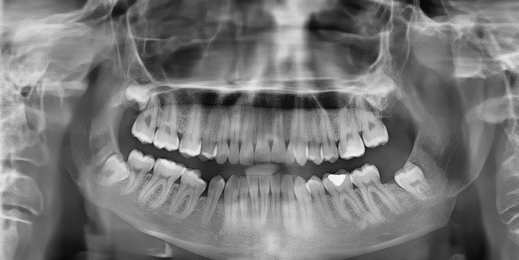} &
        \includegraphics[width=0.16\linewidth]{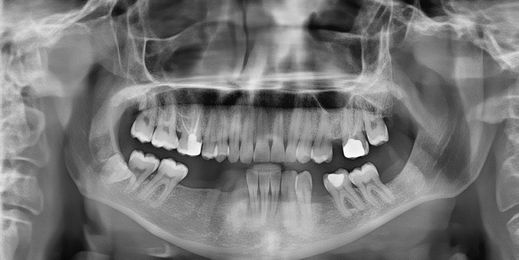} &
        \includegraphics[width=0.16\linewidth]{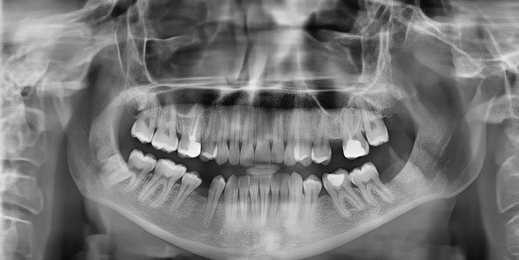} &
        \includegraphics[width=0.16\linewidth]{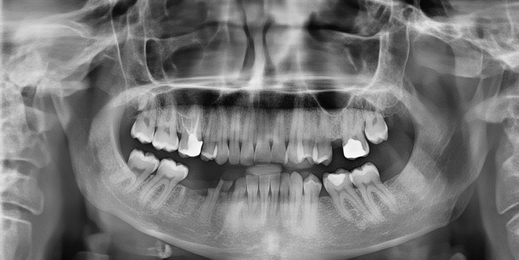} &
        \includegraphics[width=0.16\linewidth]{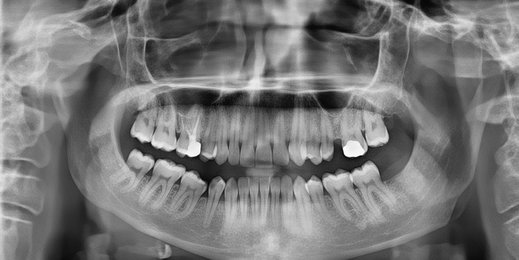} \\
        
        \includegraphics[width=0.16\linewidth]{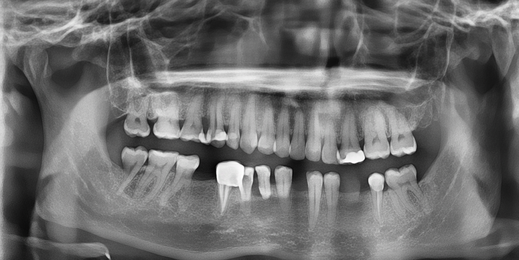} &
        \includegraphics[width=0.16\linewidth]{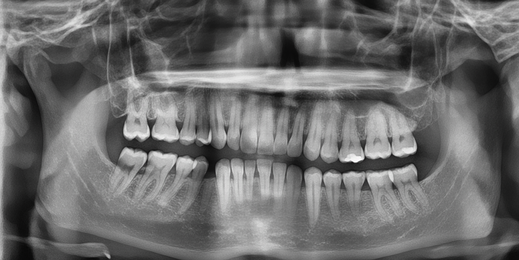} &
        \includegraphics[width=0.16\linewidth]{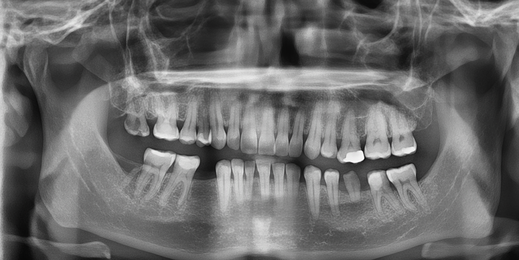} &
        \includegraphics[width=0.16\linewidth]{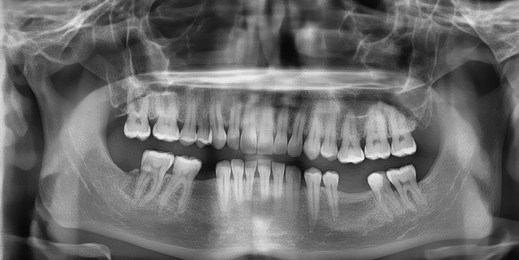} &
        \includegraphics[width=0.16\linewidth]{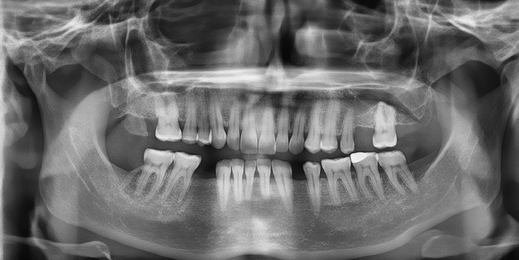} &
        \includegraphics[width=0.16\linewidth]{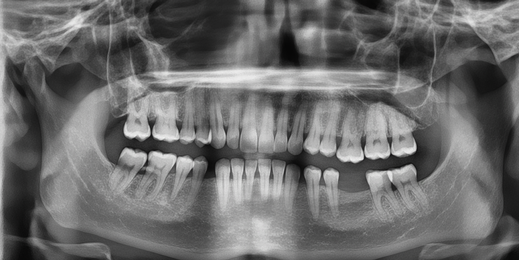} \\
        
        \includegraphics[width=0.16\linewidth]{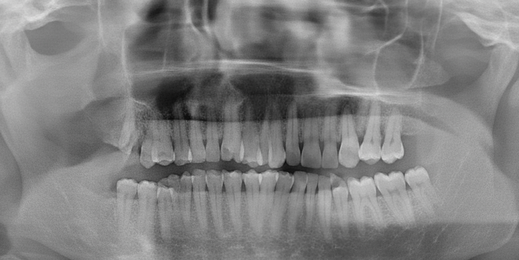} &
        \includegraphics[width=0.16\linewidth]{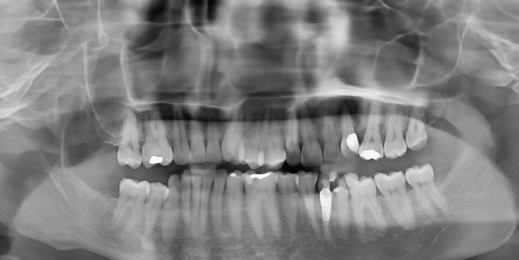} &
        \includegraphics[width=0.16\linewidth]{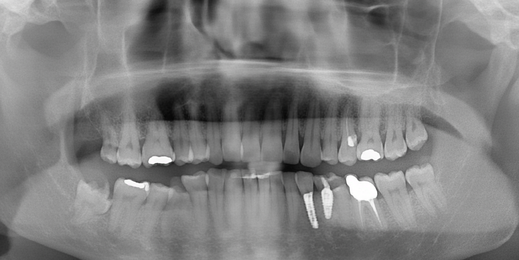} &
        \includegraphics[width=0.16\linewidth]{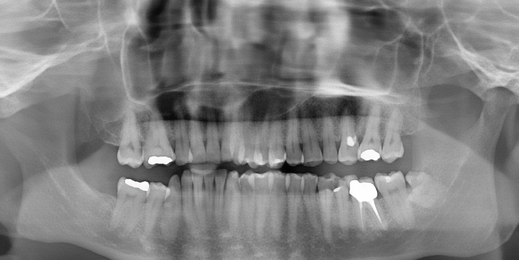} &
        \includegraphics[width=0.16\linewidth]{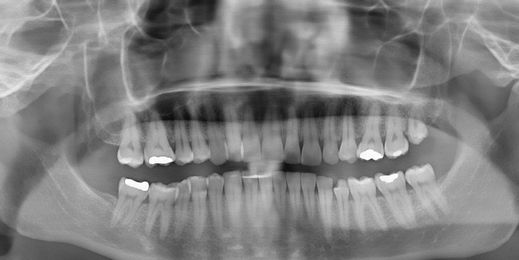} &
        \includegraphics[width=0.16\linewidth]{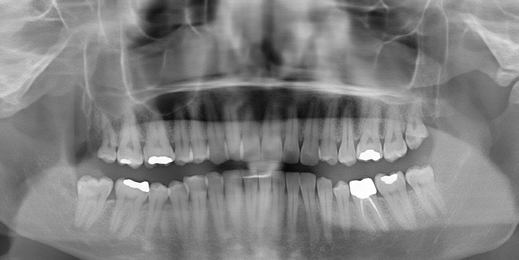} \\
        
        \includegraphics[width=0.16\linewidth]{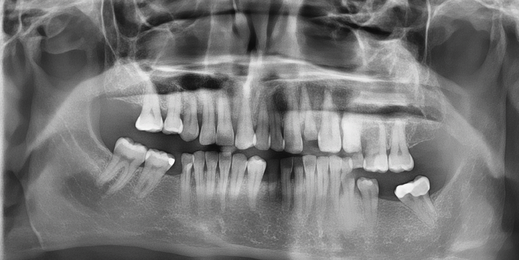} &
        \includegraphics[width=0.16\linewidth]{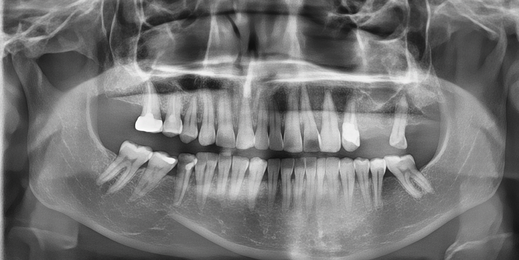} &
        \includegraphics[width=0.16\linewidth]{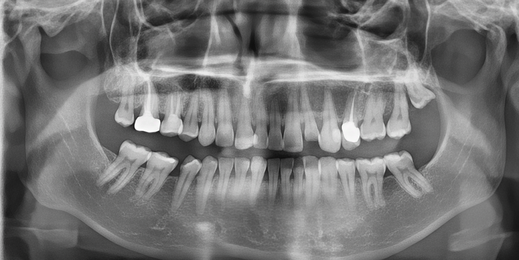} &
        \includegraphics[width=0.16\linewidth]{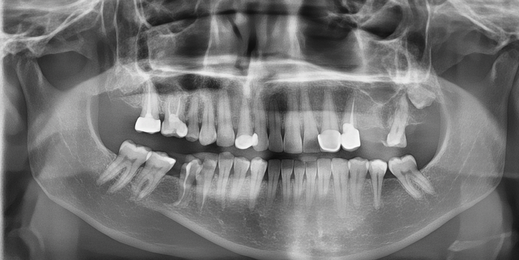} &
        \includegraphics[width=0.16\linewidth]{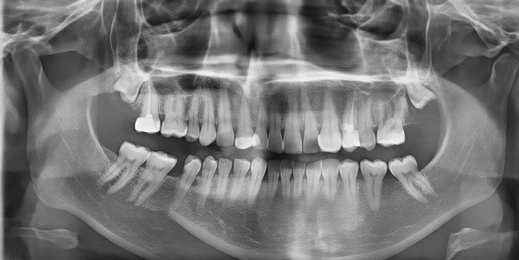} &
        \includegraphics[width=0.16\linewidth]{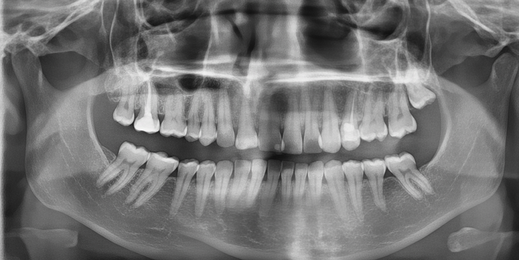} \\
    \end{tabular}
    \caption{Comparison of generated PRs across epochs. Each column represents a different epoch from left to right, showing the images generated using same unique seed per row.}
    \label{fig:epoch_comparison}
\end{figure*}

\section{Results \& Discussions}
\begin{figure*}[tp]
    \centering
    \includegraphics[width=\linewidth]{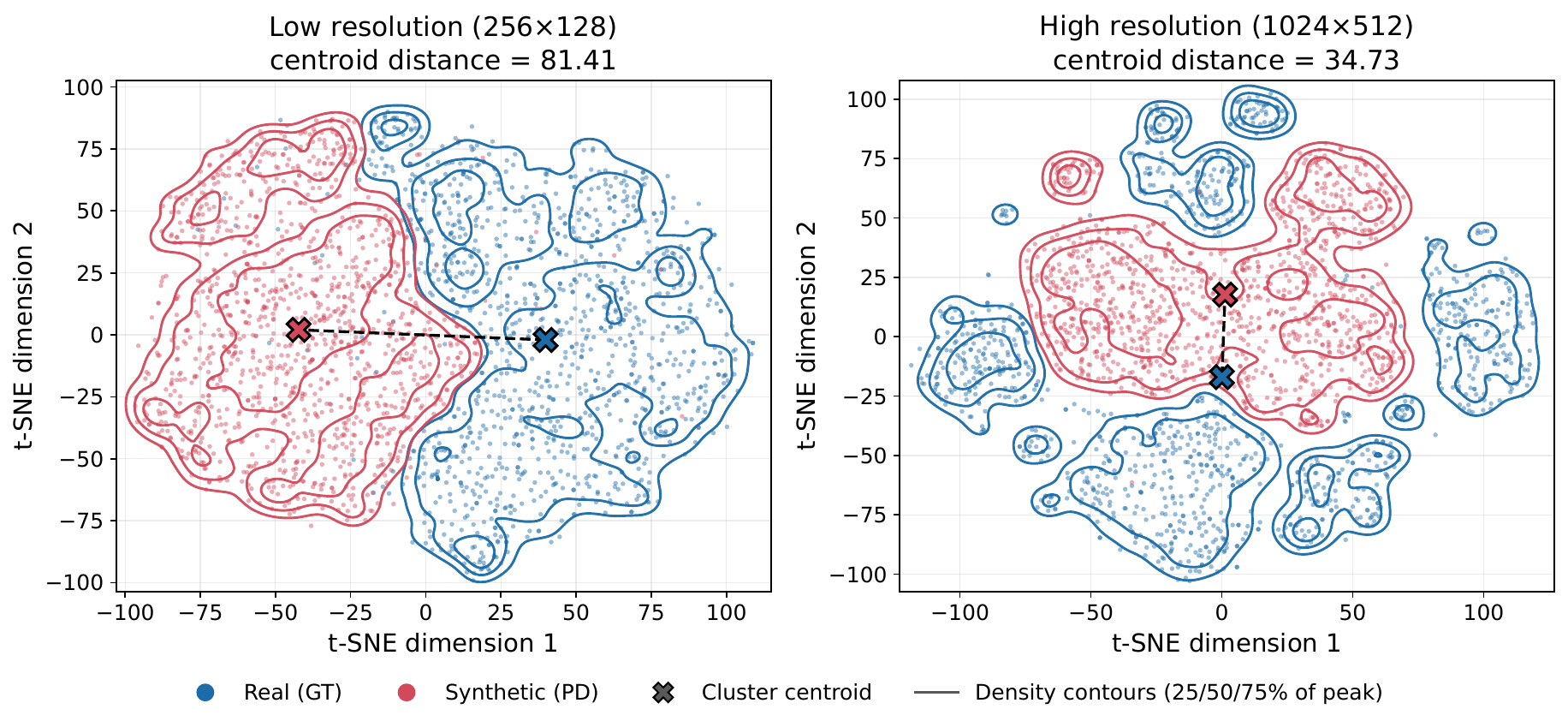}
    \caption{Visualization of t-SNE embeddings in 2D for low-resolution (LR, left) and high-resolution (HR, right) data for ground truth (GT, i.e. real data) and PanoDiff (PD, i.e. synthetic data). Euclidean distances between the centroids of these clusters are - LR: 81.41, HR: 34.73. To keep the plot readable with $2 \times 7243$ points per panel, a random subsample of 1200 points per class is drawn as a scatter, while the contours give the kernel density of the complete set at 25\%, 50\% and 75\% of its peak; crosses mark the centroids of the complete sets and the dashed line connects them.}
    \label{fig:tsne}
\end{figure*}

We define four sets of images to compare, namely (1) LR\_GT for Low-resolution Ground-truth (i.e. real PRs), (2) HR\_GT for High-resolution Ground-truth, (3) LR\_PD for Low-resolution PanoDiff (i.e. synthetic PRs), and (4) HR\_PD for High-resolution PanoDiff (after applying the SR model). Each group contains $7243$ PRs. A t-SNE plot of features extracted using ResNet-50 was created to visualize the distribution of real and synthetic PRs, shown in Figure \ref{fig:tsne}. Interestingly, while the plot shows distinct clusters for low-resolution data (although adjacent, and slightly overlapping), the high-resolution data shows t-SNE clusters for synthetic data surrounded by clusters corresponding to real data, along with a reduced Euclidean distance between the cluster centroids, which could indicate that the SR model improves the overall realism. 
Furthermore, we report FID scores for various combinations of image sources in Table \ref{tab:fidcompared}. It can be seen that HRGT-HRPD has an FID of $40.5$ indicating good overall performance of PanoDiff. We also report FID of inter-dataset real images to show the diversity of training images for PanoDiff. For combinations with \{LR, HR\} + \{GT, PD\} (first row of Table \ref{tab:fidcompared}), we split the original set into two halves and combine these with random transformations, resulting in $7243$ images for each half for FID calculation.

\begin{table*}[tp]
    \centering
    \caption{Comparison of real vs. synthetic images using FID. Lower scores indicate greater similarity.}
    \label{tab:fidcompared}
    \begin{tabular}{|l|c|c|}
        \hline
        \textbf{Comparison Type} & \textbf{Combination} & \textbf{FID} \\
        \hline
        \textbf{Self-comparisons} & HRGT-HRGT & 5.7 \\
        & LRGT-LRGT & 6.2 \\
        & HRPD-HRPD & 5.0 \\
        & LRPD-LRPD & 5.9 \\
        \hline
        \textbf{Real vs. Synthetic} & LRGT-LRPD & 55.0 \\
        & HRGT-HRPD & 40.5 \\
        & HRGT-LRPD & 108.8 \\
        & LRGT-HRPD & 94.3 \\
        \hline
        \textbf{Real vs. Real} & ADLD-DENTEX & 50.0 \\
        & ADLD-TSXK & 63.0 \\
        & ADLD-TUFTS & 36.8 \\
        & ADLD-USPFORP & 70.0 \\
        & DENTEX-TSXK & 59.1 \\
        & DENTEX-TUFTS & 38.1 \\
        & DENTEX-USPFORP & 70.1 \\
        & TSXK-TUFTS & 37.2 \\
        & TSXK-USPFORP & 60.7 \\
        & TUFTS-USPFORP & 45.5 \\
        \hline
        \textbf{HR vs. LR} & HRGT-LRGT & 53.5 \\
        & HRPD-LRPD & 112.0 \\
        \hline
    \end{tabular}
\end{table*}

\begin{table*}[tp]
    \centering
    \caption{IS, with higher values indicating more diversity and image quality. The left side of the table shows IS for \{LR, HR\} + \{GT, PD\}, the right side shows IS for the datasets of real images used in this study. G corresponds to pure Gaussian noise images.}
    \vspace{10pt}
    \begin{tabular}{lcc|lcc}
        \toprule
        Source & IS Mean & IS Std & Source & IS Mean & IS Std \\
        \midrule
        HRGT   & 2.684 & 0.064 & ADLD     & 2.938 & 0.037 \\
        HRPD   & 2.381 & 0.031 & DENTEX   & 2.905 & 0.051 \\
        LRGT   & 2.898 & 0.042 & TSXK     & 2.644 & 0.053 \\
        LRPD   & 2.983 & 0.066 & TUFTS    & 2.627 & 0.047 \\
        \quad G & 1.050 & 0.001 & USPFORP  & 2.882 & 0.034 \\
        \bottomrule
    \end{tabular}
    \label{tab:inception_scores}
\end{table*}

IS values are reported in Table \ref{tab:inception_scores}. Apart from IS for real and synthetic datasets, we also show IS for 7243 Gaussian noisy images ($\mu = 128$, $\sigma = 64$) as negative controls. Compared with HRPD (2.38), LRPD (2.98) performed better when compared with real data (GT). This suggests that the LRPD images have slightly more diversity or are perceived as more classifiable compared to the LRGT. 

\subsection{Expert Evaluation}
\label{sec:expert}
We performed an assessment of the realism of the generated images using 200 images (100 real + 100 fake) with time-limited observers, where the PRs are shown for 12 seconds (including ca. 2 seconds of loading time for each PR). Six observers were involved: three early-career (EC) dentists with less than 10 years of postgraduate clinical experience (with expertise in prosthodontics, oral surgery and orthodontics) and three experienced (EP) dentists with over 15 years of clinical expertise (with expertise in oral radiology, oral surgery and orthodontics). The observations were performed independently; each observer evaluated the images in a single session of ca. 1 hour (incl. breaks). For each PR, the observers were required to choose a value in a slider as follows: 0 for definitely real, 0.25 for probably real, 0.5 for unsure, 0.75 for probably fake and 1 for definitely fake. The observation results of the six dentists are shown in Table \ref{tab:user_metrics}.
A sample of the radiographs used in this task, with the six observers' responses to
each of them, can be inspected on the project page at
\url{https://s4nyam.github.io/panodiff/}, which also lets the reader attempt the task under
the same time limit. 

\begin{table*}[tp]
\centering
\caption{Observer responses and derived performance metrics for the six dentists.
The two left-hand blocks give the \emph{raw} number of responses in each category of the
five-level scale, separately for the 100 synthetic and the 100 real PRs; every row of each
block sums to 100 and no weighting of any kind has been applied. DR: definitely real;
PR: probably real; U: unsure; PF: probably fake; DF: definitely fake. The right-hand block
gives precision, recall and accuracy computed from the certainty-weighted counts as
described in the text.}
\label{tab:user_metrics}
\small
\setlength{\tabcolsep}{3.4pt}
\begin{tabular}{lccccc ccccc ccc}
\toprule
 & \multicolumn{5}{c}{100 synthetic PRs} & \multicolumn{5}{c}{100 real PRs} & \multicolumn{3}{c}{Metrics} \\
\cmidrule(lr){2-6} \cmidrule(lr){7-11} \cmidrule(lr){12-14}
TLD & DR & PR & U & PF & DF & DR & PR & U & PF & DF & P & R & A \\
\midrule
EC1 &  3 &  7 & 16 & 34 & 40 & 13 & 20 & 32 & 25 & 10 & 0.60 & 0.75 & 0.63 \\
EC2 &  4 & 22 &  3 & 25 & 46 & 26 & 46 &  3 & 19 &  6 & 0.68 & 0.72 & 0.69 \\
EC3 &  0 &  4 & 11 & 45 & 40 &  0 & 38 & 33 & 28 &  1 & 0.63 & 0.80 & 0.66 \\
\midrule
EP1 &  0 &  6 &  7 & 83 &  4 &  1 & 54 & 33 & 12 &  0 & 0.65 & 0.71 & 0.66 \\
EP2 &  2 & 12 &  0 & 33 & 53 & 40 & 44 &  0 & 16 &  0 & 0.78 & 0.81 & 0.79 \\
EP3 &  3 & 14 &  5 & 35 & 43 & 17 & 39 & 14 & 26 &  4 & 0.65 & 0.75 & 0.68 \\
\bottomrule
\end{tabular}
\end{table*}

\begin{figure*}[tp]
    \centering
    \includegraphics[width=\linewidth]{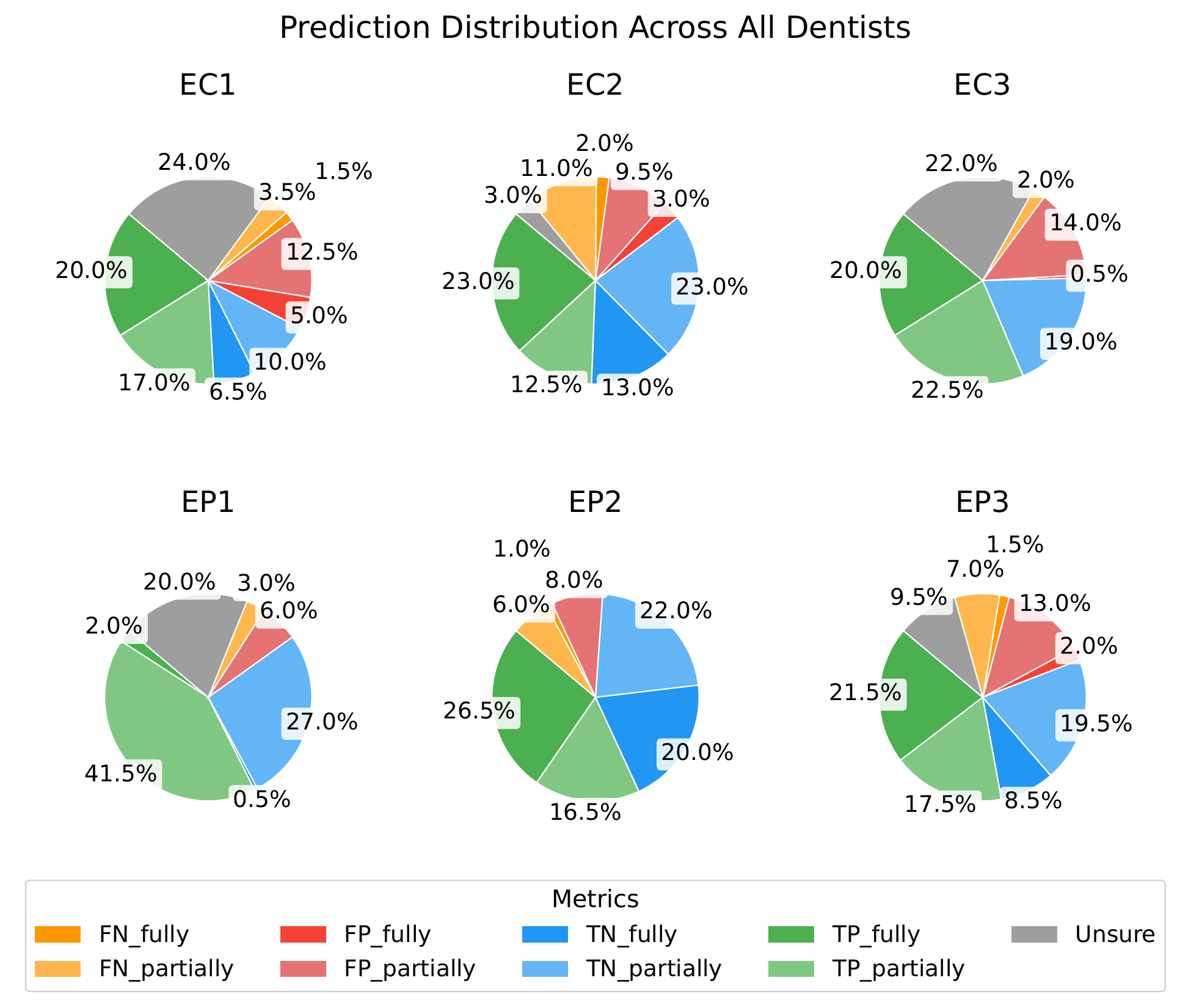}
    \caption{Pie charts for each observer showing distribution of correct and incorrect decisions. `Fully' and `partially' refers to the level of certainty indicated by the observer for a given answer, as described in the text.}
    \label{fig:piecharts}
\end{figure*}

\begin{figure*}[tp]
    \centering
    \includegraphics[width=0.75\linewidth]{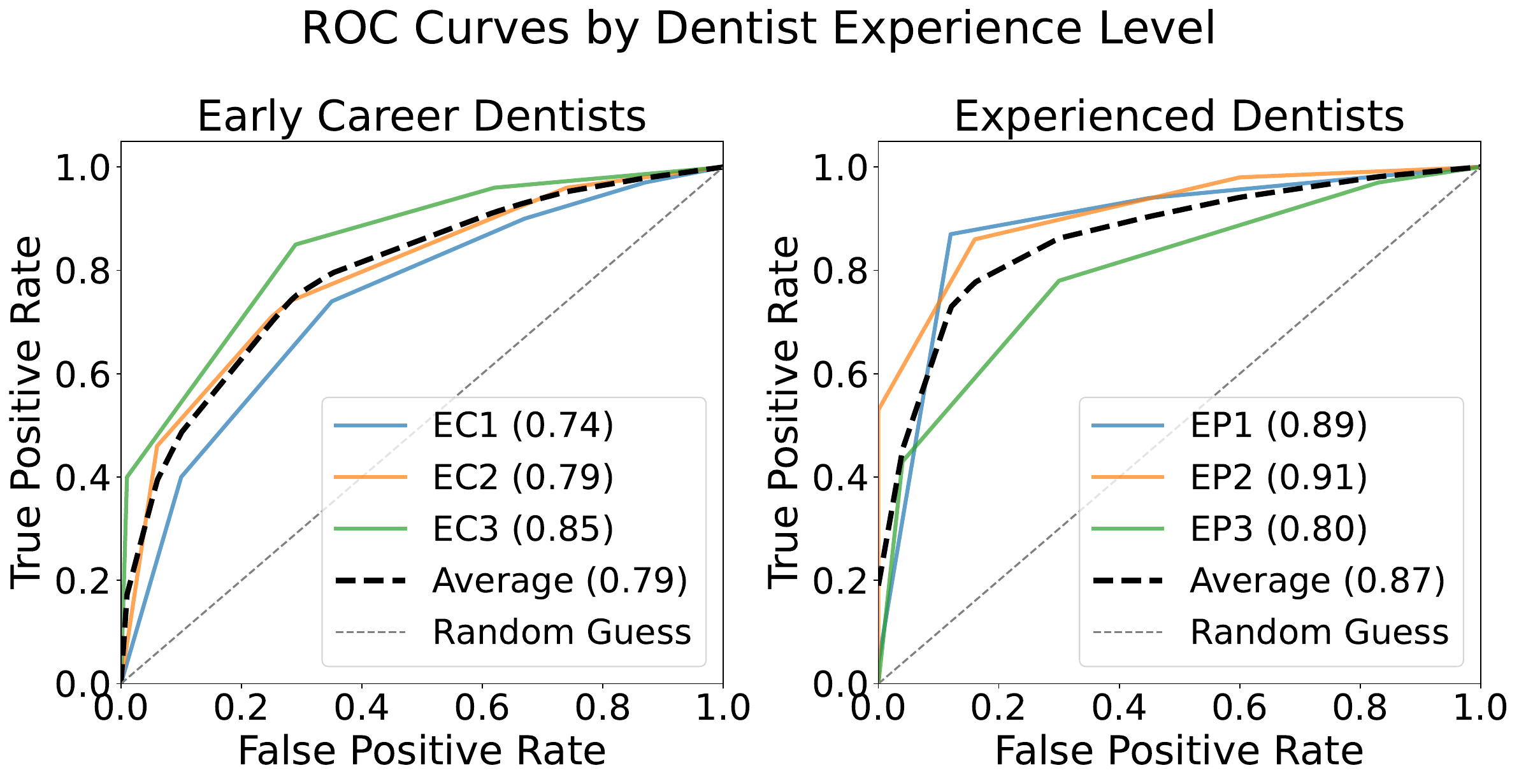}
    \caption{Left: ROC plot for individual ECs and their average. Right: ROC plot for individual EPs and their average. The value in brackets after each label is the AUC.}
    \label{fig:ecep}
\end{figure*}

\begin{figure*}[tp]
    \centering
    \includegraphics[width=0.75\linewidth]{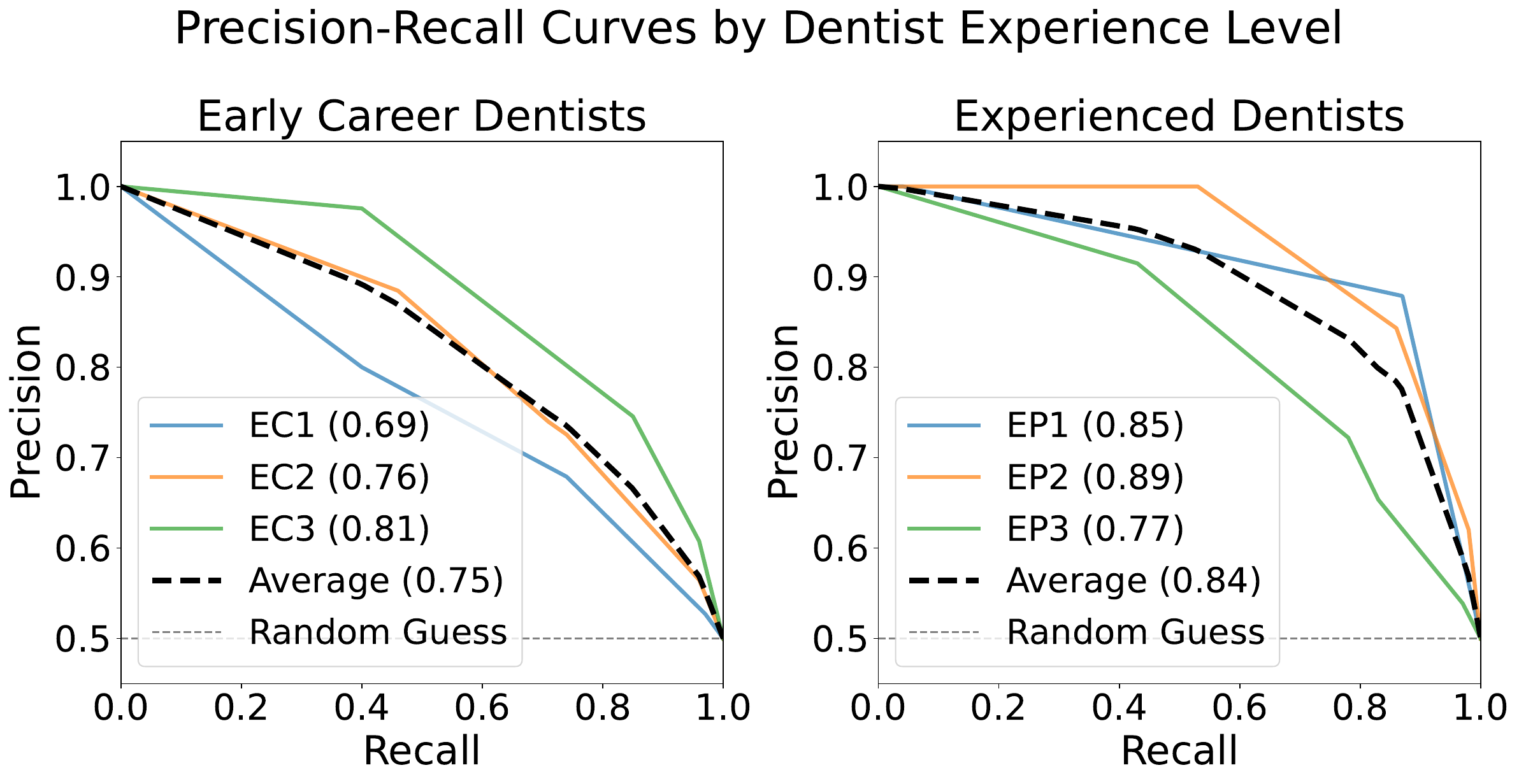}
    \caption{Left: precision--recall curve for individual ECs and their average. Right: precision--recall curve for individual EPs and their average. The value in brackets after each label is the average precision (AP).}
    \label{fig:roc-pc}
\end{figure*}

Table~\ref{tab:user_metrics} reports the observations in two forms. The two left-hand blocks are the raw response tallies: for each observer, how many of the 100 synthetic and how many of the 100 real PRs were placed in each of the five categories. The metrics in the right-hand block are derived from these responses through the intermediate quantities (a) True Positive (TP): The original image was `Fake' and was correctly predicted as `Fake', (b) True Negative (TN): The original image was `Real' and was correctly predicted as `Real', (c) False Negative (FN): The original image was `Fake' but was incorrectly predicted as `Real', (d) False Positive (FP): The original image was `Real' but was incorrectly predicted as `Fake', (e) Unsure: An image rated `Unsure' counts 0.5 towards each of the two outcomes for its original label (TN \& FP if `Real' or TP \& FN if `Fake'), (f) Precision (P): Out of all the images predicted as `Fake', Precision tells us how many were actually `Fake', calculated as TP / (TP + FP), (g) Recall (R): Out of all the images that were truly `Fake,' Recall tells us how many were correctly identified as `Fake.' calculated as TP / (TP + FN), (h) Accuracy (A): The overall correctness of the user, calculated as (TP + TN) / (TP + TN + FP + FN). For all of these metrics, the certainty of the observer was taken into account; for example, a `probably fake' response for a synthetic image would count as 0.75 times a TP and 0.25 times a FN. Observer-wise distributions of metrics, distinguishing between decisions with partial and full certainty, are shown in Figure \ref{fig:piecharts}. We show receiver operating characteristic (ROC) plots and precision--recall curves for EC and EP dentists, along with the average for each group, in Figure \ref{fig:ecep} and Figure \ref{fig:roc-pc}, respectively. The ROC curve shows a higher average area under the curve (AUC) for EPs (0.87) than for ECs (0.79) at distinguishing real from fake PRs.

Because the graded scale can be scored in different ways, we report two alternatives. Weighting an unsure response as half a correct and half an incorrect decision, which
is the convention used above and throughout the paper, yields a mean accuracy of $68.4\%$
($68.5\%$ when averaged from the two-decimal values in Table~\ref{tab:user_metrics}).
Counting each ``probably'' response as a full decision instead, with each unsure response either counted as an error or resolved to ``fake'', gives $68.3\%$ and $71.8\%$ respectively. The three conventions therefore agree to within
$3.5$ percentage points.

We analyzed the observers' results further. Each reader evaluated 100 real and 100 synthetic PRs in the same session, so the real images serve as the control condition and a reader who would guess randomly would score 50\%. Three questions follow. First, is 68.5\% better than random guessing? Treating each reader as one observation, the six accuracies give a 95\% confidence interval of 62.6--74.3\%, that is, the range in which the true mean accuracy of readers of this kind would be expected to lie; it excludes 50\%. A one-sample $t$-test, which asks how many standard errors the observed mean sits away from a reference value, places it $8.15$ standard errors above chance with five degrees of freedom ($p=0.0005$), and the exact Wilcoxon signed-rank test, a rank-based alternative that makes no assumption of normality and so suits a sample of only six readers, agrees ($p=0.031$, the smallest value attainable with six observations). Individually, every reader's area under the ROC curve exceeds 0.5 by DeLong's test, which compares AUCs using the variance of the underlying rank statistic (all $p<10^{-11}$). We therefore do not claim the synthetic images are undetectable. Second, is 68.5\% lower than what observers achieve on GAN-generated PRs? This question can be partially answered by comparing our results with previously published values using one-sample $t$-tests of our six reader accuracies against the published values, showing significant differences with the 78.2\% sensitivity with which five clinicians identified full-size StyleGAN2-ADA radiographs \cite{schoenhof2024synthetic} ($p=0.008$), and the 82.3\% accuracy with which readers identified StyleGAN3 radiographs containing jaw cysts \cite{fukuda2024attempt} ($p=0.002$). Neither study is a like-for-like replication of ours (the viewing protocols, reader panels and image sets differ), so the tests should be read as comparisons against published reference points rather than as a controlled head-to-head trial. The task was also difficult in the other direction: the certainty-weighted responses counted 38.9\% of the real radiographs as synthetic. Third, do experienced observers perform better at distinguishing real from fake radiographs? That depends on what is treated as the sample. Across images, the group-consensus ROC curves separate clearly (AUC 0.94 for EPs against 0.86 for ECs; DeLong $z=3.75$, $p=0.0002$). With three readers per group, however, the difference between the readers themselves is not significant (exact permutation test, $p=0.20$), so we report the advantage of the experienced readers as consistent across the images these readers saw, but not established for dentists in general.

\begin{figure*}[tp]
\centering
\begin{tabular}{|c|c|c|c|c|}
\hline
 & \multicolumn{2}{c|}{\textbf{TP}} & \multicolumn{2}{c|}{\textbf{FN}} \\
\hline
\raisebox{3.5\height}{\textbf{FC}} & 
\includegraphics[width=0.2\textwidth]{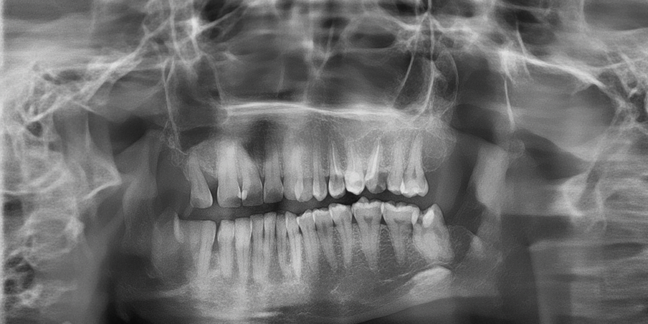} & 
\includegraphics[width=0.2\textwidth]{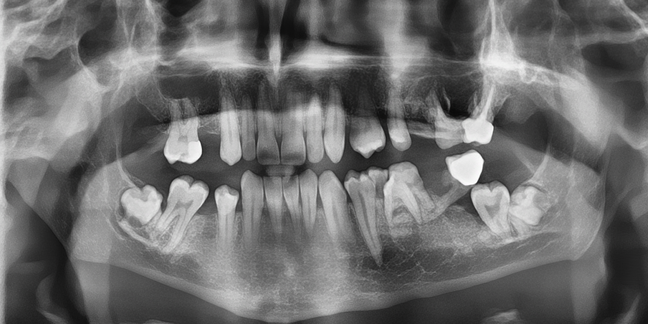} & 
\includegraphics[width=0.2\textwidth]{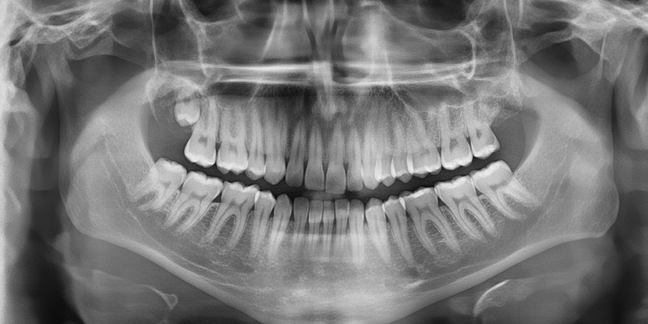} & 
\includegraphics[width=0.2\textwidth]{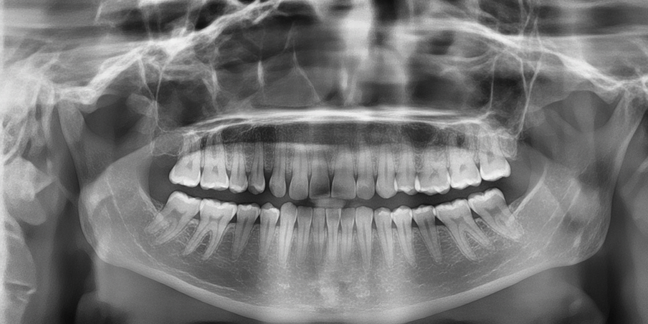} \\
\hline
\raisebox{3.5\height}{\textbf{PC}} & 
\includegraphics[width=0.2\textwidth]{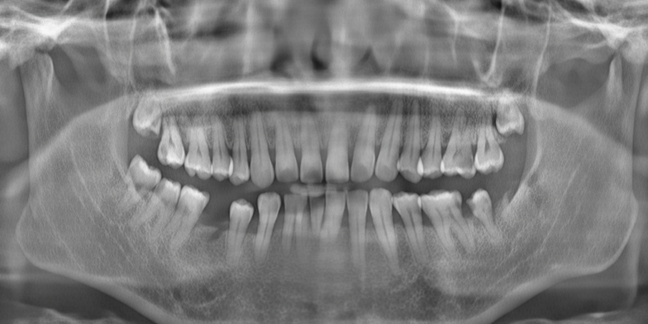} & 
\includegraphics[width=0.2\textwidth]{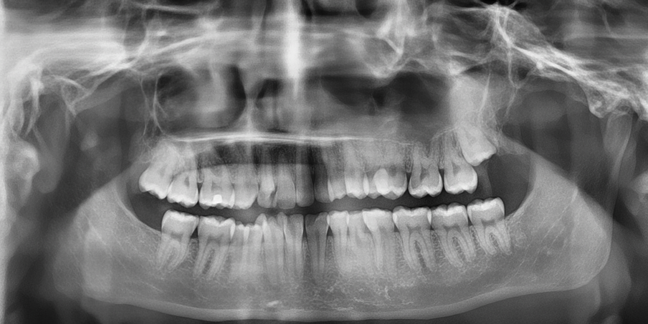} & 
\includegraphics[width=0.2\textwidth]{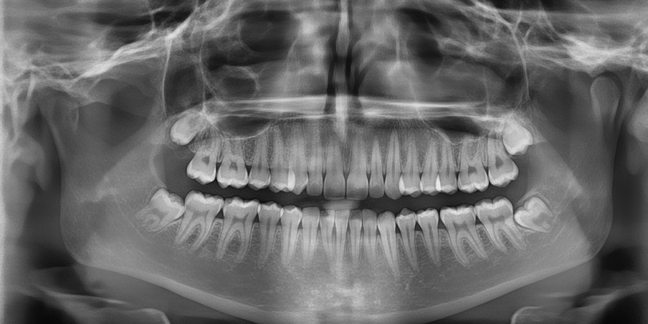} & 
\includegraphics[width=0.2\textwidth]{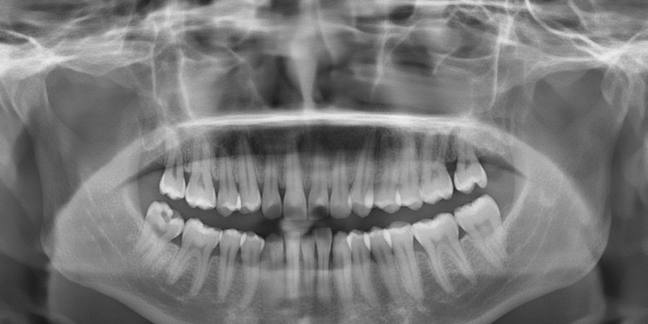} \\
\hline
 & \multicolumn{2}{c|}{\textbf{FP}} & \multicolumn{2}{c|}{\textbf{TN}} \\
\hline
\raisebox{3.5\height}{\textbf{FC}} & 
\includegraphics[width=0.2\textwidth]{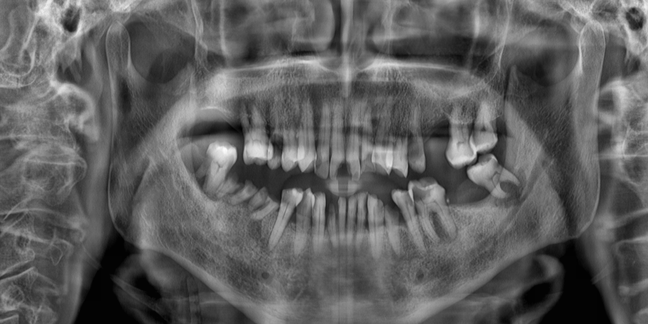} & 
\includegraphics[width=0.2\textwidth]{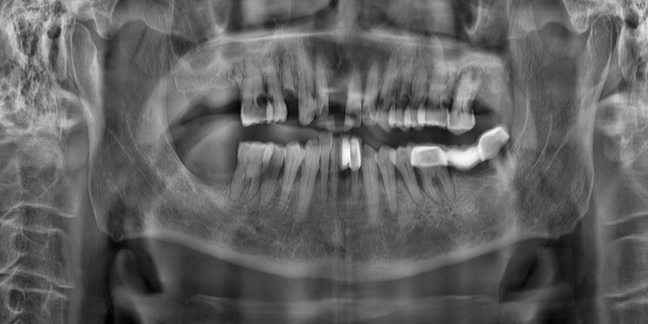} & 
\includegraphics[width=0.2\textwidth]{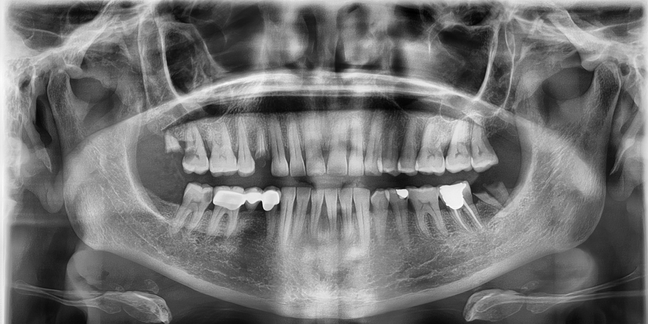} & 
\includegraphics[width=0.2\textwidth]{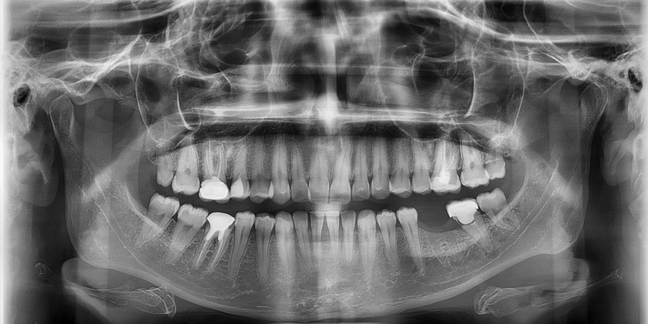} \\
\hline
\raisebox{3.5\height}{\textbf{PC}} & 
\includegraphics[width=0.2\textwidth]{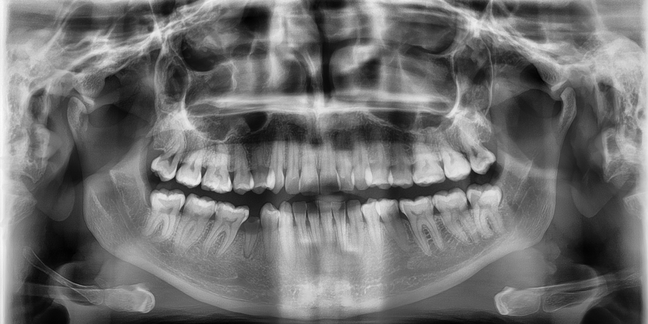} & 
\includegraphics[width=0.2\textwidth]{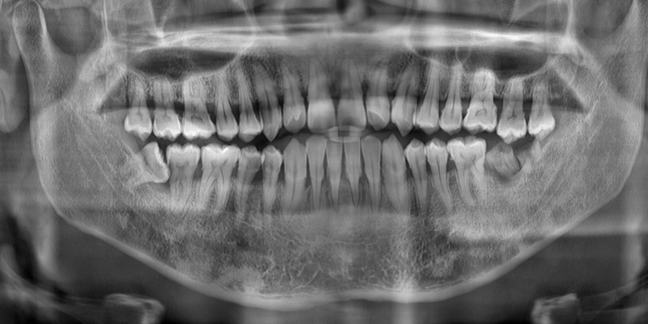} & 
\includegraphics[width=0.2\textwidth]{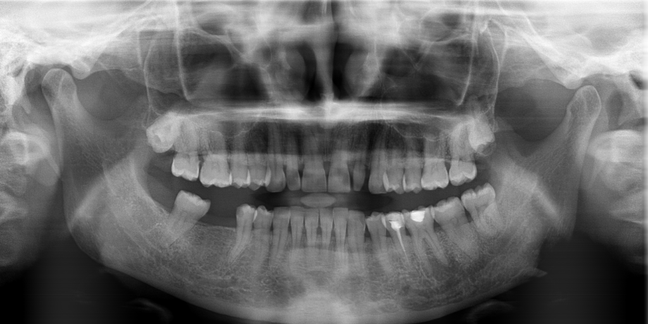} & 
\includegraphics[width=0.2\textwidth]{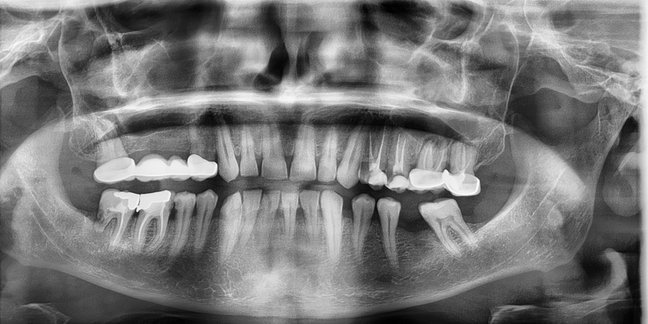} \\
\hline
\end{tabular}
\caption{Examples of PRs from expert evaluation. Within each category, the first row corresponds to fully certain (FC) and the second row to partially certain (PC) responses. Examples were selected for each category based on the majority of the observers' assessments. }
\label{fig:confusion_matrix_xrays}
\end{figure*}

Figure \ref{fig:confusion_matrix_xrays} illustrates a few TP, FN, FP and TN examples from the real-vs-fake observation. 

The images used in the observation also inherit the imperfections of the training data. Metal
objects, mild positioning errors and moderate distortion were deliberately retained
(Section~\ref{sec:datasets}) because such findings are routine rather than exceptional in
clinical panoramic radiography; positioning errors alone affected $89\%$ of $1782$
consecutive radiographs in one study \cite{dhillon2012positioning}, and are the most
frequent error type reported \cite{rondon2014common}. Because PanoDiff samples
unconditionally from that pooled distribution, it reproduces these phenomena rather than
suppressing them: radiopaque objects, asymmetries of the
occlusal plane consistent with head tilt, and the widened or narrowed ramus appearance
associated with anteroposterior mispositioning all recur among the synthetic PRs, as do
crowns, fillings and root canal fillings at plausible locations. Having such features is particularly useful if the images are to be pooled with real data or used to show trainees what
imperfect radiographs look like. One difference from real images is visible on close
inspection, and it is among the cues available to a clinician under time-limited viewing:
metal objects in a real PR have a definite physical signature, saturating the detector and producing a sharply defined `real image' and, when they lie outside the image layer, casting a characteristic `ghost image' elsewhere in the radiograph with more ill-defined borders \cite{mcdavid1983real}, whereas the synthetic counterparts are
regions of high intensity whose borders are smoother.

\subsection{Inter-observer Agreement using Cohen's Kappa}

To assess the inter-observer agreement, Cohen's kappa statistic ($\kappa$) was calculated for all pairs of observers. Cohen's kappa measures the degree of agreement between two raters, with a value of $1$ indicating perfect agreement and a value of $0$ corresponding to random agreement.

We used a \emph{weighted kappa} with quadratic weights, which takes into account the ordinal nature of the scores. Weighted kappa applies penalties based on the severity of disagreement between raters:  
\begin{itemize}
    \item Small disagreements (e.g., ``definitely real'' vs. ``probably real'') are penalized less.
    \item Large disagreements (e.g., ``definitely real'' vs. ``definitely fake'') are penalized more heavily.
\end{itemize}

The results are presented in Table~\ref{tab:kappa_weighted}. Values are qualitatively interpreted according to the scale by Landis and Koch \cite{kappalandis}.
\begin{table*}[tp]
\centering
\caption{Weighted Cohen's Kappa values between observers.}
\label{tab:kappa_weighted}
\begin{tabular}{lcccccc}
\toprule
 & EC1 & EC2 & EC3 & EP1 & EP2 & EP3 \\
\midrule
EC1 & 1 & 0.42 & 0.50 & 0.44 & 0.47 & 0.53 \\
EC2 & 0.42 & 1 & 0.56 & 0.45 & 0.49 & 0.49 \\
EC3 & 0.50 & 0.56 & 1 & 0.53 & 0.50 & 0.62 \\
EP1 & 0.44 & 0.45 & 0.53 & 1 & 0.57 & 0.47 \\
EP2 & 0.47 & 0.49 & 0.50 & 0.57 & 1 & 0.50 \\
EP3 & 0.53 & 0.49 & 0.62 & 0.47 & 0.50 & 1 \\
\bottomrule
\end{tabular}
\end{table*}

Overall, the observed kappa values ranged between \textbf{0.42 and 0.62}, indicating that the interobserver agreement in this dataset was \emph{moderate}, with one pair of observers (EC3 and EP3) showing \emph{substantial} agreement. There was no meaningful difference in agreement between early-career and experienced dentists. Moderate rather than high agreement is consistent with the difficulty observers faced in reliably distinguishing real from synthetic images.

Three observations clarify why the agreement, while clearly above chance, is not higher in a real-versus-synthetic discrimination task.

The first concerns the graded response scale. Rather than forcing a binary decision, observers rated each PR on a five-level scale spanning ``definitely real'', ``probably real'', ``unsure'', ``probably fake'' and ``definitely fake'', so that partial certainty could be expressed directly. The distribution of responses over this scale is itself a result. As the two ``U'' columns of Table~\ref{tab:user_metrics} show, several observers placed a substantial share of their responses at the midpoint (48 for EC1, 44 for EC3 and 40 for EP1, out of 200), which is a direct statement by practising clinicians that, for roughly a fifth to a quarter of the images, no evidence favouring either category was available to them. Other observers preferred to commit: EP2 in every case and EC2 in all but six. Both are legitimate uses of a graded scale, and the freedom to differ is precisely what the scale was designed to permit. Because weighted $\kappa$ is computed relative to each rater's own marginal distribution, this variation in response style lowers the coefficient independently of whether the raters actually order the images similarly.

The second concerns the reading conditions, which were designed to be favourable to the observers. All six were practising dentists who interpret PRs routinely, and judging whether a radiograph is anatomically and physically plausible falls squarely within that competence: implausible arch geometry, inconsistent radiopacity or unnatural texture are the kinds of deviation clinicians are trained to notice. The 12-second display window must also be seen in proportion to the judgement being asked for, which is one of realism and not of diagnosis. Holistic impressions of an image form very quickly: natural scenes can be categorised reliably from presentations of about 150~ms \cite{thorpe1996speed}, and radiologists discriminate abnormal from normal mammograms at above-chance rates after exposures of only 250--500~ms, before any systematic search is possible \cite{evans2013gist}. Twelve seconds is one to two orders of magnitude beyond the exposure at which such global impressions become available, and is therefore ample for anatomically or physically implausible content to register, even though it is deliberately shorter than the time a full diagnostic reading would take. That readers so equipped nevertheless converged only moderately on which images were synthetic is therefore informative about the images rather than about the observers: the cues that would have supported consistent identification were present in only part of the images.

The third is the statistical consequence of that difficulty. When a substantial subset of images carries little discriminative signal, responses on that subset approach independent guessing, and $\kappa$, which measures agreement in excess of chance, necessarily falls, however skilled the raters are. The moderate $\kappa$ and the 68.5\% mean accuracy are thus two views of a single finding: the readers shared some cues, enough to agree beyond chance and to score above it, but a sizeable fraction of the synthetic PRs were not reliably separable from real ones.

\subsection{Attention Maps}
\label{sec:attention}
\begin{figure*}[tp]
    \centering
    \includegraphics[width=\linewidth]{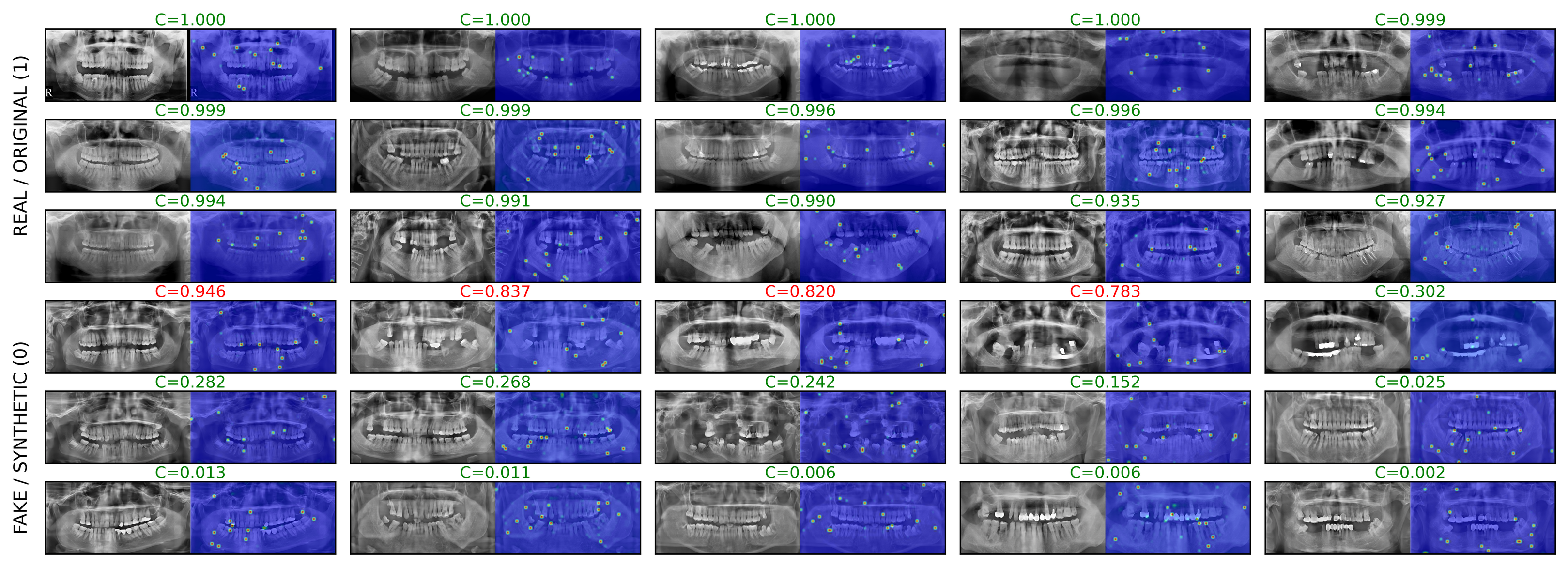}
    \caption{Attention maps generated using a trained ViT for two classes - top three rows for real images and bottom three rows for synthetic images. The confidence value `C' represents the ViT classifier's output and ranges from 0 (for synthetic) to 1 (for real). Prediction values in red correspond to incorrect classification.}
    \label{fig:attentionmaps}
\end{figure*}

In order to get a visual perspective regarding the performance of our synthetic PR generator, we trained a custom ViT for distinguishing real from synthetic PRs on $7243$ images ($1024 \times 512$) of HR\_GT (class 0) and an equal number of images from HR\_PD (class 1). The ViT showed a $97.5\%$ test accuracy with a train-test split of 80:20, drawn over distinct radiographs so that no radiograph repeated in the corpus (Section~\ref{sec:datasets}) falls on both sides (mean of four training runs). Next, we visualized the attention maps (AMs). We postulate that AMs will approximate the important regions that contribute to the predicted label for the PRs. Figure \ref{fig:attentionmaps} shows 30 random samples from the test data with AMs. It can be observed that for real images, the model consistently focused on anatomically significant regions such as the teeth, mandibular area, and mouth structures. In contrast, for synthetic images, the attention appeared more diffuse. Interestingly, in some synthetic images, especially those where the diffusion model achieved high realism, the ViT’s AMs began to resemble those seen in real images (red labels in Figure \ref{fig:attentionmaps}). This points toward a range of realism in the synthetic images, where certain synthetic examples closely mimic true anatomical features. Such observations serve as an implicit evaluation of the diffusion model’s generative quality.

The contrast between the two groups of maps admits a concrete interpretation. On real PRs, the attention concentrates on regions that are dense in fine, quasi-periodic structure: the dentition and its restorations, the alveolar crest, and the inferior border of the mandible. These are precisely the regions where a real radiograph carries high-frequency detail that is highly constrained by physics and anatomy (trabecular bone pattern, the enamel--dentin boundary, the periodontal ligament space), and where the joint statistics of neighbouring pixels are hardest to imitate. On synthetic PRs the attention is more diffuse. Read together, these two observations suggest that the classifier is not recognising anatomy as such, but is instead detecting whether the expected fine-scale texture is present and correctly localised: on synthetic images no region satisfies that test as decisively, and the attention disperses. This is consistent with the fact that our generator produces the image in two stages, since the $256 \times 128$ diffusion stage fixes the global arch geometry, which the maps indicate is reproduced convincingly, whereas all detail below that scale is reconstructed by the SR stage from a prior learned on natural images and fine-tuned on PRs, and is plausible rather than physically derived.

The cases in which a synthetic PR draws a real-like attention pattern (red labels in Figure~\ref{fig:attentionmaps}) fit the same account: these are samples for which the SR stage happened to produce locally coherent dental texture, and they are also the samples most likely to deceive a human observer. This points to a range of realism within the synthetic set rather than a uniform quality level, and it suggests that per-image quality control, rather than a single dataset-level FID, is what a downstream user of these images would actually need.

Two cautions apply to this analysis. First, the gap between the ViT's 97.5\% test accuracy and the observers' 68.5\% mean accuracy shows that a trained detector exploits systematic low-level signatures that clinicians do not perceive under time-limited viewing. Realism with respect to a human reader and realism with respect to a learned detector are therefore different properties, and our observer results should not be taken to imply the latter. This distinction is practically important, because a downstream model trained on these images is more like the ViT than like the dentist. Second, attention weights are a visualisation of where the model reads, not a faithful attribution of what determined its output, and the limits of treating attention as explanation are well documented \cite{jain2019attention}. The observations above are accordingly based on a qualitative reading of 30 test samples, so we measured them over the full test set as well.

\begin{figure*}[tp]
\centering
\includegraphics[width=\linewidth]{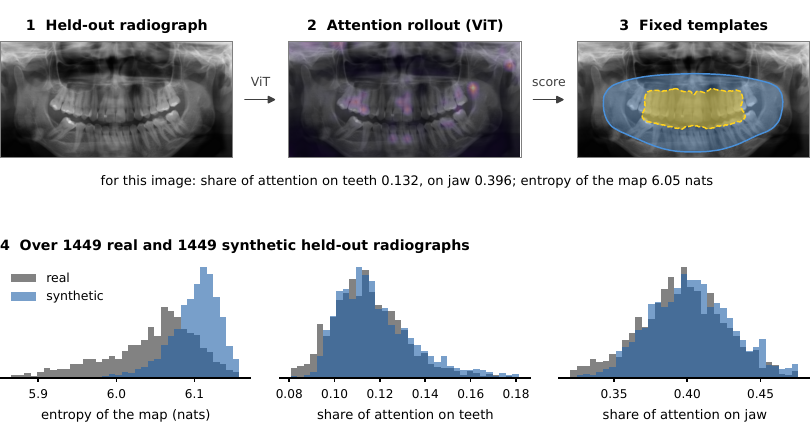}
\caption{Attention statistics. Top: a held-out radiograph (1), the corresponding
attention-rollout map from the real-versus-synthetic ViT (2), and the fixed
anatomical templates (3), comprising the TUFTS teeth masks (dashed) and maxillomandibular masks (solid), averaged and thresholded at $0.5$. For every image, the share of attention
inside the teeth template, inside the jaw template, and the entropy of the map were calculated. Bottom (4):
distributions of those numbers over the $1449$ real and $1449$ synthetic held-out radiographs.
The entropy separates the two classes, whereas the anatomical shares overlap considerably.}
\label{fig:attnprocess}
\end{figure*}

To that end, attention-rollout maps \cite{abnar2020quantifying} of a ViT with the same architecture and training recipe were computed
for all $2898$ held-out radiographs ($1449$ real, $1449$ synthetic). Each map was scored against
fixed anatomical templates, the TUFTS expert masks of the teeth (the 968 that are not empty) and of the maxillomandibular region (all 1000), each averaged and thresholded at $0.5$, so that the regions are identical for every image
(Figure~\ref{fig:attnprocess}): the fraction of attention mass falling on the teeth and on the
maxillomandibular region, together with the entropy of the map as a scale-free measure of how
concentrated it is. The measurement bears out the diffuseness seen in the samples. Each difference is tested with the Mann--Whitney $U$ test, and its size is given by the rank-biserial correlation $r=2\,\mathrm{AUC}-1$, where the AUC is the probability that a randomly chosen synthetic image scores higher than a real one: $r=0$ means the two classes are indistinguishable and $r=1$ that they are fully separated.
Attention on synthetic radiographs is more spread out than on real ones (entropy $6.100 \pm 0.033$ against $6.044 \pm 0.062$ nats), and this is by a wide margin the
largest effect in the analysis ($r=+0.607$, i.e.\ an AUC of $0.80$ and a difference of $1.1$ pooled standard deviations; $p<10^{-175}$ because each class contains $1449$ images). The
share of attention falling on anatomy, by contrast, is nearly the same for both classes:
$0.119$ against $0.115$ on the teeth and $0.403$ against $0.396$ on the jaw, with small
effect sizes ($r=+0.107$ and $+0.124$, AUCs of $0.55$ and $0.56$) that reach significance only because each group
contains $1449$ images. What distinguishes the two classes over the full test set is
therefore how concentrated the attention is, rather than where it falls.

\subsection{Probing with a medical segmentation foundation model}
\label{sec:medsam}
In this section, we assess whether a pretrained segmentation model treats a synthetic radiograph the same way it treats a real one. To this end, MedSAM \cite{ma2024segment} was used, which is a general medical segmentation model, fine-tuned from the Segment Anything Model (SAM) on over 1.5 million image--mask pairs from various imaging modalities and organs (not containing any real or synthetic images used in the present study). Given a box around an object, it returns a mask of that object and a score for how good it believes the mask to be. If synthetic teeth look like real teeth to such a model, it should segment them with the same confidence and accuracy. This follows Isola et al.~\cite{isola2017image}, who judged generated images by how well a segmentation network trained on real images segments them.

We used all 7243 synthetic radiographs and the 2697 real radiographs that have a manual tooth
mask, that is, a segmentation of the teeth drawn by hand by the annotators of each dataset. The manual masks are needed to verify MedSAM's ability to segment teeth and to train a model that identifies the tooth region. The latter model used a U-Net with a ResNet-50 encoder, trained on a 70\% split (1888) of the manual masks and applied in the same way to the real and the synthetic radiographs. The real radiographs come from four of the datasets of Table~\ref{tab:datasetoverview}: all
500 ADLD radiographs, which have a polygon for every tooth; the 634 DENTEX radiographs of its
quadrant-enumeration subset, the only part of DENTEX with tooth outlines (the rest is
unlabelled or carries quadrant or diagnosis labels only); the 598 TSXK radiographs that come
with a mask; and the 1000 TUFTS radiographs with expert teeth masks. USPFORP has bounding boxes
only. Three TSXK and 32 TUFTS masks are empty, so those radiographs were dropped, leaving $500 + 634 + 598 + 1000 - 35 = 2697$ radiographs. All were preprocessed in the same way as the training corpus (Section~\ref{sec:datasets}), with each manual mask cropped and resized with its radiograph. 

Every image received the same two kinds of box prompt: a \emph{fixed} box at the same
position in every image, so that no model of ours is involved, and a \emph{dentition} box around the tooth region determined by the aforementioned U-Net. For each prompt
we recorded MedSAM's own quality score, the size and shape of its mask (area, number of parts, and extent, the share of its bounding box that it fills) and the Dice overlap between its mask and the U-Net tooth region. On real images we also computed the Dice with the manual mask. We report the effect size as the rank-biserial $r$ of Section~\ref{sec:attention}, where
$|r|<0.1$ is negligible, $0.1$--$0.3$ small and above $0.3$ medium. To give these values a scale, the same $r$ was computed between each real
device and the other three.

\begin{table*}[tp]
\centering
\caption{MedSAM under identical box prompts on 2697 real and 7243 synthetic radiographs. Predicted quality is MedSAM's own score for its mask;
Dice is the overlap between MedSAM's mask and the U-Net tooth region. $r$ is the effect
size of the real--synthetic difference ($r>0$: higher on real; $|r|<0.1$ negligible,
$0.1$--$0.3$ small, $>0.3$ medium) and $p$ is the two-sided Mann--Whitney value. The last column
gives, for scale, the range of $|r|$ when each real device is compared with the other three on
the same measure.}
\label{tab:medsam}
\adjustbox{max width=\linewidth}{\begin{tabular}{llccccc}
\toprule
\textbf{Prompt} & \textbf{Measure} & \textbf{Real} & \textbf{Synthetic} & $\boldsymbol{r}$ & $\boldsymbol{p}$ & \textbf{Device vs.\ others} $|r|$ \\
\midrule
Fixed box & predicted quality & $0.464$ & $0.466$ & $-0.02$ & $0.20$ & $0.09$--$0.32$ \\
          & Dice to reference & $0.571$ & $0.563$ & $+0.01$ & $0.34$ & $0.02$--$0.32$ \\
Dentition box & predicted quality & $0.480$ & $0.472$ & $+0.08$ & $3 \times 10^{-10}$ & $0.07$--$0.23$ \\
              & Dice to reference & $0.794$ & $0.788$ & $+0.03$ & $0.04$ & $0.01$--$0.19$ \\
\bottomrule
\end{tabular}}
\end{table*}

\begin{figure*}[tp]
\centering
\includegraphics[width=\linewidth]{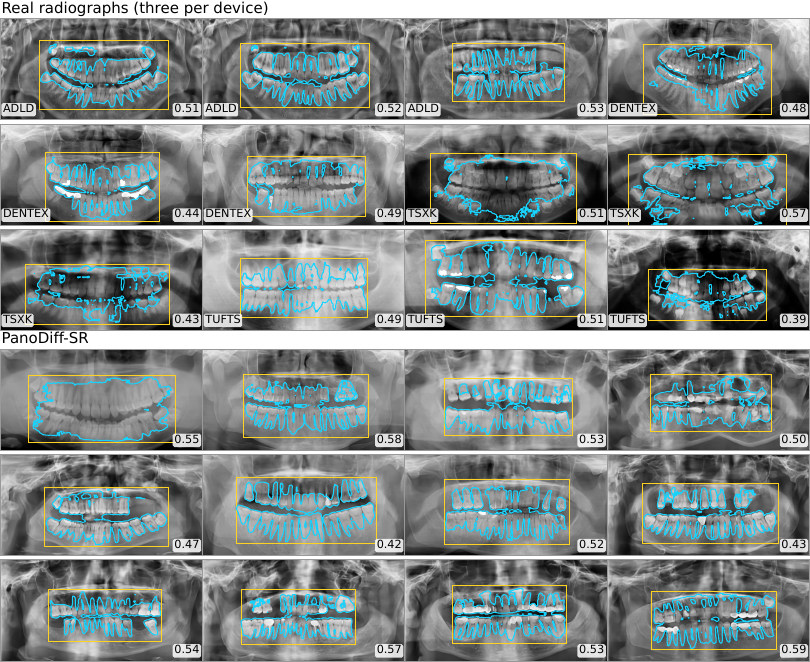}
\caption{MedSAM with the dentition-box prompt on real radiographs, three from each labelled device, and on PanoDiff-SR radiographs. The yellow box is the prompt, drawn around the U-Net tooth region in the same way on both sets; the cyan contour is MedSAM's mask; the badge is its predicted mask quality.}
\label{fig:medsam}
\end{figure*}

Table~\ref{tab:medsam} gives the results and Figure~\ref{fig:medsam} shows examples. First, it can be seen that MedSAM successfully segments teeth on panoramic radiographs: its Dice score against the manual masks is $0.78$ with the dentition box. Second, MedSAM treats the two sets almost the same under both prompts. Its quality score is $0.464$ against $0.466$ with the fixed box and $0.480$ against $0.472$ with the dentition box. Both differences are negligible ($|r| \le 0.08$). The Dice with the reference region and the size and shape of the masks differ only slightly ($|r| \le 0.11$). Third, the real devices differ from each other by more than the difference between real and synthetic images: comparing
each device with the other three on the same measures gives $|r|$ of up to $0.32$.
In concrete terms, a segmentation model that was trained on neither set segments the teeth of a
synthetic PR as readily as those of a real one, which indicates that the dental anatomy in the
synthetic images is realistic enough for an off-the-shelf medical imaging model to process.

\subsection{Comparison with alternative models and across source devices}
\label{sec:comparative}

This section reports three comparisons. Section~\ref{sec:lrsr} compares each stage of the pipeline
with one alternative: the diffusion generator (PanoDiff) with a GAN (FastGAN \cite{liu2021towards})
at $256 \times 128$, and HAT-SR \cite{chen2023hat} with SwinIR \cite{liang2021swinir} for upscaling
either generator's samples to $1024 \times 512$. Section~\ref{sec:directhr} compares the complete
two-stage pipeline with six generators trained directly at $1024 \times 512$.
Section~\ref{sec:controls} compares the synthetic radiographs with the real radiographs of each
source device. 

\subsubsection{PanoDiff versus FastGAN, and HAT-SR versus SwinIR}
\label{sec:lrsr}
\begin{figure*}[p]
\centering
\includegraphics[width=\linewidth,height=0.81\textheight,keepaspectratio]{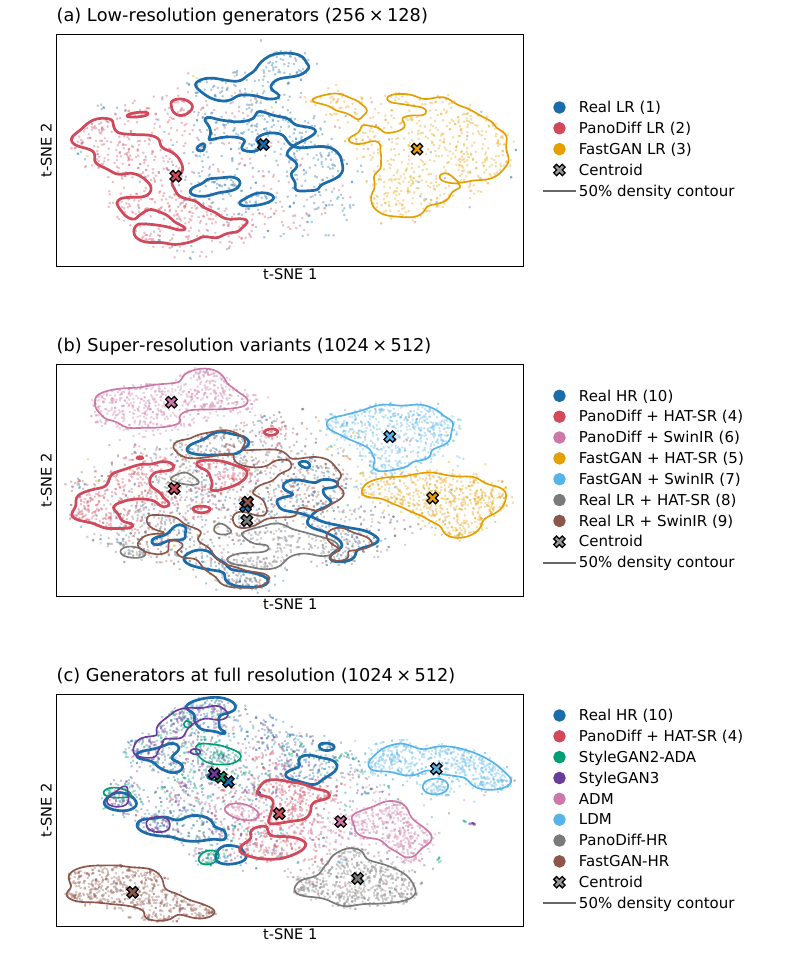}
\caption{t-SNE of every generative arm in the Inception feature space used for all FID and
IS values, on 1500 images per arm; each panel is embedded separately (PCA to 50 dimensions, then
t-SNE with perplexity 30). Numbers in parentheses are the source indices of
Table~\ref{tab:supp-inception-scores}. Crosses mark cluster centroids and lines enclose half of
each arm's peak density. Distances between clusters in a t-SNE map are not quantitatively
meaningful, but measured in the original feature space, the distance from each full-resolution
generator in panel (c) to the real images orders the arms exactly as their FIDs in
Table~\ref{tab:directhr} do.}
\label{fig:tsnearms}
\end{figure*}

In this comparison, every source (listed in Table~\ref{tab:supp-inception-scores}) contains 7243 PRs.
Both generators were trained from randomly initialised weights on the 7243 real PRs, and both
upscalers were fine-tuned on the same 7243 real PRs, each with the objective and degradations of its
original training recipe (a pixel loss on fixed fourfold downsampling for SwinIR; pixel, perceptual and
adversarial losses on varied degradations for HAT-SR). Figure~\ref{fig:tsnearms} gives an overview of
the sets compared in this section (panels a and b) and in Section~\ref{sec:directhr} (panel c), as
t-SNE maps of the Inception features on which the measures below are computed.

\begin{table*}[tp]
\centering
\caption{IS for different sources, each of 7243 images; the index is used by Table~\ref{tab:supp-FID-scores} and Figures~\ref{fig:tsnearms} and~\ref{fig:prscores}. HAT-SR and SwinIR were both fine-tuned on the 7243 real PRs. LR: low-resolution, HR: high resolution. GT: ground truth (i.e., real images), PD: PanoDiff diffusion, GAN: generative adversarial network.}

\begin{tabular}{|c|l|c|c|}
\hline
\textbf{Index} & \textbf{Source} & \textbf{IS Mean} & \textbf{IS Std} \\
\hline
1 & LRGT & 2.898 & 0.042 \\
\hline
2 & LRPD & 2.983 & 0.066 \\
3 & LRGAN & 2.253 & 0.034 \\
\hline
4 & HRPD & 2.381 & 0.031 \\
5 & LRGAN+HATSR & 2.047 & 0.020 \\
\hline
6 & LRPD+SwinIR & 2.887 & 0.073 \\
7 & LRGAN+SwinIR & 2.172 & 0.046 \\
\hline
8 & LRGT+HATSR & 2.735 & 0.068 \\
9 & LRGT+SwinIR & 2.750 & 0.053 \\
\hline
10 & HRGT & 2.684 & 0.064 \\
\hline
\end{tabular}
\label{tab:supp-inception-scores}
\end{table*}

\begin{table*}[tp]
\centering
\caption{Distances to the real set at the same resolution for each generated set. Source numbers are the indices of Table~\ref{tab:supp-inception-scores}; every set contains 7243 images. FID and KID measure the distance between the Inception feature distributions, and lower is closer; KID is reported $\times 10^3$ as mean $\pm$ sd over 100 subsets of 1000 images. Precision (fidelity) and recall (coverage) \cite{kynkaanniemi2019improved} use the same features with $k=3$ nearest neighbours, and higher is better. LR: low-resolution, HR: high resolution. GT: ground truth (i.e., real images), PD: PanoDiff diffusion, GAN: generative adversarial network.}

\adjustbox{max width=\linewidth}{\begin{tabular}{|c|l|c|c|c|c|}
\hline
\textbf{Source 1} & \textbf{Source 2} & \textbf{FID}~$\downarrow$ & \textbf{KID}~$\downarrow$ & \textbf{Precision}~$\uparrow$ & \textbf{Recall}~$\uparrow$ \\
\hline
1-LRGT & 2-LRPD & 55.0 & $60.5 \pm 2.3$ & 0.283 & 0.220 \\
1-LRGT & 3-LRGAN & 85.2 & $100.2 \pm 3.3$ & 0.429 & 0.015 \\
\hline
10-HRGT & 4-HRPD & 40.5 & $40.8 \pm 2.0$ & 0.562 & 0.288 \\
10-HRGT & 5-LRGAN+HATSR & 102.3 & $136.9 \pm 3.0$ & 0.069 & 0.008 \\
10-HRGT & 6-LRPD+SwinIR & 94.4 & $118.5 \pm 2.5$ & 0.045 & 0.026 \\
10-HRGT & 7-LRGAN+SwinIR & 109.7 & $140.2 \pm 3.6$ & 0.167 & 0.006 \\
\hline
\end{tabular}}
\label{tab:supp-FID-scores}
\end{table*}

\begin{figure*}[tp]
\centering
\includegraphics[width=\linewidth]{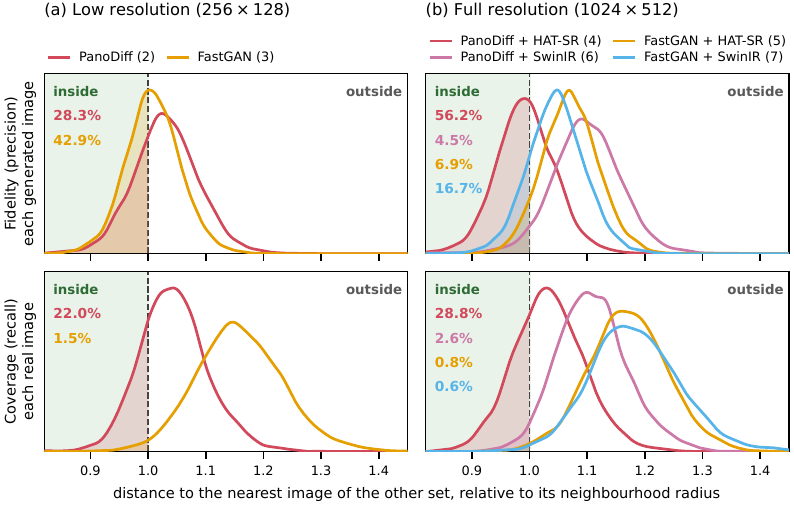}
\caption{Precision (fidelity) and recall (coverage) for (a) the two generators at low resolution and (b) the four
combinations at full resolution. Every image has a neighbourhood in the Inception feature
space: the region within the distance to its third-nearest neighbour in its own set. Top row:
for each generated image, its smallest distance to a real image, measured in units of that real
image's neighbourhood radius; images left of the dashed line lie inside the neighbourhood of at
least one real image, and their share, printed in each panel in the colour of its curve, is the
precision. Bottom row: the same for each real image with respect to the generated images; the
share inside is the recall. Curves represent kernel density estimates over all 7243 images of each
set. Numbers in parentheses are the source indices of Table~\ref{tab:supp-inception-scores}.}
\label{fig:prscores}
\end{figure*}

\begin{figure*}[p]
\centering
\includegraphics[width=\linewidth]{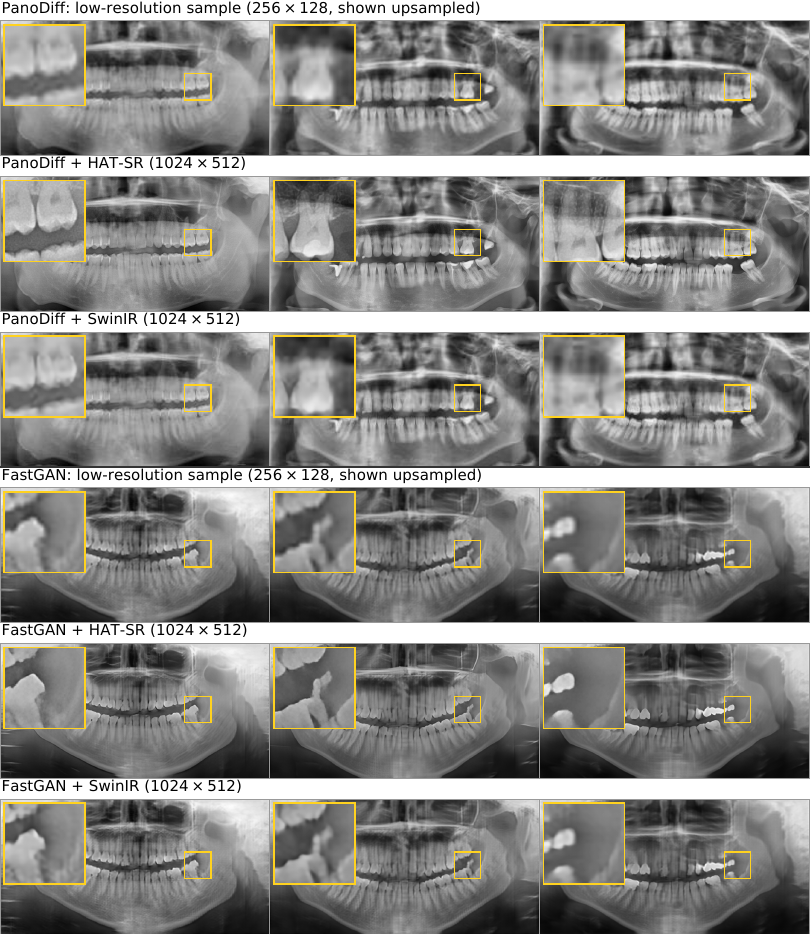}
\caption{Low-resolution synthesis and super-resolution for three samples of each
generator (columns). Each column traces one seed from the $256 \times 128$ sample, upsampled
bicubically for display, through HAT-SR and SwinIR, for PanoDiff (top three rows) and FastGAN
(bottom three rows). The inset in every panel magnifies the region outlined in yellow.}
\label{fig:lrsr}
\end{figure*}

Each generated set is compared with the real set at the same resolution, using measures
computed on the same Inception features. FID and KID
\cite{binkowski2018demystifying} measure how far apart the two sets are, and lower is better.
KID is an unbiased estimate, so it depends less than FID on the number of images. Precision and
recall \cite{kynkaanniemi2019improved} show where a distance comes from. Precision is the share
of generated images that lie close to real images in this feature space (fidelity); recall is
the share of real images that lie close to generated images (coverage).
The Inception score needs no real reference set: it is high when a network trained to classify
natural images (Inception) assigns each generated image to one class with confidence while the
set as a whole is spread over many classes \cite{borji2019pros}.

At low resolution, PanoDiff is closer to the real data than FastGAN in terms of both FID and
KID (Table~\ref{tab:supp-FID-scores}). Figure~\ref{fig:prscores}a shows that FastGAN has a higher precision than PanoDiff, but a very low recall, showing the limited
diversity (mode collapse) that adversarial training is known for.

At high resolution, PanoDiff with HAT-SR is the closest of the four combinations to the
real images on every measure: it has the lowest FID and KID and the highest precision and
recall, each by a wide margin (Table~\ref{tab:supp-FID-scores} and Figure~\ref{fig:prscores}b). HAT-SR generally outperformed SwinIR for FID and KID, with the difference being more pronounced for PanoDiff than for FastGAN, whose
low-resolution output limits what either upscaler can recover. Figure~\ref{fig:prscores}b shows the distribution of precision and recall for the different high-resolution combinations.

Figure~\ref{fig:lrsr} shows three samples of each generator, including low-resolution versions and
outputs from both upscalers. SwinIR returns an image close to a smooth enlargement of its
input, whereas HAT-SR restores fine structure such as enamel edges and trabecular bone. We
measured this with a Laplacian filter, which responds to fine detail: taking the median over
all 7243 images of the mean squared response, PanoDiff's samples after SwinIR keep about
$4\%$ of the high-frequency energy of real radiographs, against about $80\%$ after HAT-SR
(vs. $8\%$ and $60\%$, respectively, for FastGAN).

The Inception score (Table~\ref{tab:supp-inception-scores}) ranks the upscalers the other way round: SwinIR scores higher for both
generators, and
PanoDiff's samples even score above the real radiographs. That the smoothest images score
above real radiographs shows that the Inception score does not
measure radiographic realism here: it rewards images that the ImageNet-trained Inception
network confidently assigns to one of its object classes
\cite{borji2019pros,kynkaanniemi2023role}; this has limited meaning for radiographs,
and the same limitation has been reported for synthetic medical images and chest radiographs
\cite{skandarani2023gans,segal2021evaluating}. 

\subsubsection{Two-stage pipeline versus direct high-resolution generation}
\label{sec:directhr}
\begin{table*}[tp]
\centering
\caption{Direct high-resolution synthesis compared with the two-stage pipeline. Rows 1 and 2 generate at $256 \times 128$ and are upscaled to $1024 \times 512$ by HAT-SR; rows
3--8 generate $1024 \times 512$ directly and were trained from randomly initialised weights on the same 7243 real PRs
with comparable iterations and training budget. FID
and IS are computed against the same 7243 real HR PRs. ``ViT det.'' is the held-out accuracy of the real-versus-synthetic
classifier of Section~\ref{sec:attention} retrained per arm, so lower means harder to tell from
a real radiograph. Precision and recall are computed as in Table~\ref{tab:supp-FID-scores} and
Figure~\ref{fig:prscores} ($k=3$). Training cost is in
GPU-hours, all on the same type of GPU. $^{\dagger}$Measured time per training step of each stage multiplied by its
number of training steps; the SR stage accounts for $61.5$ of these hours in both rows.}
\label{tab:directhr}
\small\setlength{\tabcolsep}{3pt}\begin{tabular}{@{}clcccccc@{}}
\toprule
\textbf{\#} & \textbf{Model} & \begin{tabular}[b]{@{}c@{}}\textbf{FID}\\\textbf{$\downarrow$}\end{tabular} & \begin{tabular}[b]{@{}c@{}}\textbf{IS}\\\textbf{$\uparrow$}\end{tabular} &
\begin{tabular}[b]{@{}c@{}}\textbf{Prec.}\\\textbf{(\%)\,$\uparrow$}\end{tabular} & \begin{tabular}[b]{@{}c@{}}\textbf{Rec.}\\\textbf{(\%)\,$\uparrow$}\end{tabular} &
\begin{tabular}[b]{@{}c@{}}\textbf{ViT det.}\\\textbf{(\%)\,$\downarrow$}\end{tabular} & \begin{tabular}[b]{@{}c@{}}\textbf{Cost}\\\textbf{(GPU-h)}\end{tabular} \\
\midrule
1 & PanoDiff-SR (ours) & 40.50 & 2.38 & 56.2 & 28.8 & 97.6 & 82$^{\dagger}$ \\
2 & FastGAN + HAT-SR \cite{liu2021towards,chen2023hat} & 102.27 & 2.05 & 6.9 & 0.8 & 99.9 & 64$^{\dagger}$ \\
3 & StyleGAN2-ADA \cite{karras2020training} & 23.45 & 2.58 & 61.1 & 21.9 & 89.6 & 441 \\
4 & StyleGAN3 \cite{karras2021alias} & 19.38 & 2.66 & 67.3 & 25.5 & 77.9 & 572 \\
5 & FastGAN-HR \cite{liu2021towards} & 215.89 & 2.30 & 0.0 & 0.0 & 100.0 & 11 \\
6 & ADM \cite{dhariwal2021diffusion} & 65.03 & 2.71 & 28.2 & 16.6 & 98.4 & 122 \\
7 & LDM \cite{muller2023multimodal} & 112.96 & 2.14 & 21.3 & 3.0 & 100.0 & 62 \\
8 & PanoDiff-HR (ours) & 155.41 & 2.24 & 27.5 & 1.7 & 100.0 & 120 \\
\bottomrule
\end{tabular}
\end{table*}

\begin{figure*}[p]
\centering
\includegraphics[width=\linewidth,height=0.82\textheight,keepaspectratio]{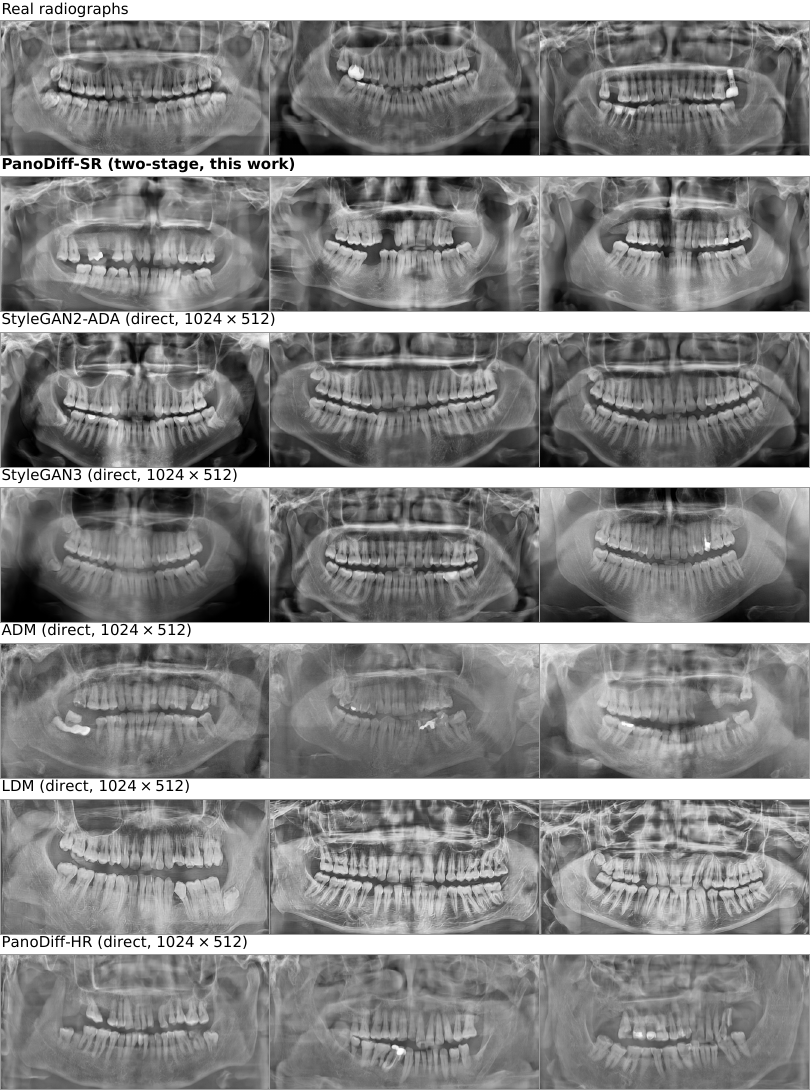}
\caption{Selected examples of the image sets of Table~\ref{tab:anatomy}: real radiographs, the
two-stage pipeline (PanoDiff-SR) and the direct high-resolution models of Table~\ref{tab:directhr}.}
\label{fig:directhr}
\end{figure*}

In this section, models that generate the
full-resolution image directly are compared using the same metrics as the previous section. Six
generators were trained from randomly initialised weights on the same 7243 real PRs at $1024 \times 512$ under a
training budget comparable to PanoDiff's, so that every model sees the data equally often. These models are compared with the proposed two-stage (generation and upsampling) PanoDiff-SR.
\textbf{PanoDiff-HR} is a high-resolution variant of PanoDiff in which the U-Net was widened to operate
natively at full resolution, \textbf{ADM}
\cite{dhariwal2021diffusion} uses pixel-space diffusion, the \textbf{latent diffusion model
(LDM)} follows Medfusion \cite{muller2023multimodal}, \textbf{StyleGAN2-ADA}
\cite{karras2020training} was used in prior work on synthetic PRs \cite{schoenhof2024synthetic}, \textbf{StyleGAN3} \cite{karras2021alias} is
its alias-free successor, which has been used for PRs with jaw cysts \cite{fukuda2024attempt},
and \textbf{FastGAN-HR} is a FastGAN variant \cite{liu2021towards} trained to yield the
full-resolution image directly.
Table~\ref{tab:directhr} reports FID, IS, precision and recall against the same 7243 real HR
PRs, the detection accuracy of the classifier of Section~\ref{sec:attention} retrained per arm,
and training cost; Figure~\ref{fig:directhr} shows selected examples of all models except the two based on FastGAN, and
Figure~\ref{fig:tsnearms}c shows the six direct high-resolution models and PanoDiff-SR in the Inception feature space.

The results show the value of the two-stage pipeline. PanoDiff-HR is markedly worse than the proposed pipeline, with samples that
are visibly blurred and low-contrast; the other direct diffusion models, ADM and LDM,
also fall behind the two-stage pipeline. The GAN results confirm this finding, with 
FastGAN followed by HAT-SR being about twice as close to the real data as the same generator
trained to yield $1024 \times 512$ images directly under an identical budget. In both families,
then, confining generation to a small grid and delegating resolution to a single
feed-forward pass produces better images at this data scale.

StyleGAN2-ADA and StyleGAN3 showed lower FIDs than PanoDiff-SR ($23.45$ and $19.38$ against $40.50$); furthermore, the classifier of Section~\ref{sec:attention} finds them harder to
detect ($89.6\%$ and $77.9\%$ against $97.6\%$). They are also the most expensive models in
the table to train, at $441$ and $572$ GPU-hours against $82$ for PanoDiff-SR. It can also be seen that the advantage of the GAN models lies in fidelity, not in coverage. Computed image by image as in
Figure~\ref{fig:prscores} on all 7243 images of each model, the two StyleGANs place more of their images close to real
radiographs (precision $61.1\%$ and $67.3\%$ against $56.2\%$), but PanoDiff-SR covers more of
the real radiographs than any other model in Table~\ref{tab:directhr}: its recall is $28.8\%$,
against $21.9\%$ for StyleGAN2-ADA and $25.5\%$ for StyleGAN3 (paired McNemar tests over the
same 7243 real images, $p<10^{-6}$). The same holds when the comparison is repeated on ten random
subsets of 5000 generated images per model: the recall (mean $\pm$ standard deviation) is
$29.6 \pm 0.7\%$ for PanoDiff-SR, against $24.4 \pm 0.5\%$ for StyleGAN2-ADA ($p<0.001$, Welch's
$t$-test) and $28.6 \pm 0.7\%$ for StyleGAN3 ($p=0.005$), and the precision is $56.1 \pm 0.3\%$,
against $60.9 \pm 0.3\%$ and $67.3 \pm 0.3\%$. PanoDiff-SR therefore reproduces more of the variety of real PRs,
and the lower FID of the GANs comes from their fidelity rather than from their diversity.
None of the models copied its training images: in the Inception feature space, the median distance
from a generated PR to its nearest training image exceeded the median distance between a real PR
and its nearest real neighbour for every model in Table~\ref{tab:directhr}. 

Additional, domain-specific metrics were defined and calculated to supplement the more conventional metrics. Notably, FID compares pooled feature statistics
over a set and is insensitive to whether any individual radiograph is anatomically coherent;
low FID has been found to coexist with anatomical implausibility severe enough that
implausible samples were removed by hand before expert review \cite{barr2026synthetic}, and
reduced FID has been shown to indicate distributional similarity rather than clinical
validity \cite{wang2026detectability}. To address this limitation, three properties every real PR has and a generator can
fail to reproduce were defined and measured automatically on 1500 images per model and for the real images: bilateral symmetry of the
jaw region, periodicity of the tooth row along the arch, and the depth of the occlusal separation (Table~\ref{tab:anatomy}). All three are computed from pixel intensities in fixed regions, built from the TUFTS expert masks: the tooth region, made up of the pixels marked as tooth in more than $35\%$ of the 968 non-empty teeth masks, and the jaw region, made up of the pixels marked as jaw in more than half of the 1000 maxillomandibular masks. Symmetry is defined as the Pearson correlation between the pixels of the jaw region and the pixels at the mirrored position on the contralateral side. Periodicity is the share of the power spectrum of the column-mean intensity profile of the tooth band (the rectangle enclosing the tooth region) that lies at periods of 20 to 60 pixels, which represents the spacing of individual teeth at $1024 \times 512$. The occlusal separation is how far the darkest row in the middle third of the tooth band falls below the fainter of the two tooth rows (the brightest rows in its top and bottom quarters), relative to the range of the band's row-mean intensity profile. Outlier rates were defined as the fraction of images outside the range of the central 95\% of real
radiographs (the 2.5th to 97.5th percentile, the range commonly used for reference intervals) on any measure
(last column of Table~\ref{tab:anatomy}); the models rank in the same order with the central 98\% or 99\%.
Similar to the previous findings, StyleGAN2-ADA, StyleGAN3 and PanoDiff-SR match the real images the most,
while ADM, LDM and particularly PanoDiff-HR show worse performance. 

\begin{table*}[tp]
\centering
\caption{Anatomical plausibility measures, on 1500 images per model. The three measures are
given as mean $\pm$ standard deviation; $^{*}$ and $^{**}$ mark a difference from the real images at
$p<0.05$ and $p<10^{-4}$ (Mann--Whitney $U$). The last column is the fraction of images falling outside
the interval spanned by the central 95\% of real radiographs on any of the three measures,
so the first row shows what a generator that matches the real distribution would score.}
\label{tab:anatomy}
\small \begin{tabular}{lcccc}
\toprule
 & \textbf{Bilateral} & \textbf{Tooth-row} & \textbf{Occlusal} & \textbf{Outside} \\
\textbf{Images} & \textbf{symmetry $r$} & \textbf{periodicity} & \textbf{separation} & \textbf{real range} \\
\midrule
Real & $0.477 \pm 0.199$ & $0.133 \pm 0.116$ & $0.503 \pm 0.283$ & $14.4\%$ \\
\midrule
PanoDiff-SR & $0.276 \pm 0.172^{**}$ & $0.169 \pm 0.124^{**}$ & $0.571 \pm 0.259^{**}$ & $20.1\%$ \\
StyleGAN2-ADA & $0.562 \pm 0.188^{**}$ & $0.113 \pm 0.108^{**}$ & $0.498 \pm 0.272$ & $16.2\%$ \\
StyleGAN3 & $0.491 \pm 0.195^{*}$ & $0.111 \pm 0.113^{**}$ & $0.500 \pm 0.278$ & $17.0\%$ \\
ADM & $0.178 \pm 0.145^{**}$ & $0.161 \pm 0.110^{**}$ & $0.475 \pm 0.275^{*}$ & $27.1\%$ \\
LDM & $0.204 \pm 0.126^{**}$ & $0.241 \pm 0.137^{**}$ & $0.596 \pm 0.262^{**}$ & $26.3\%$ \\
PanoDiff-HR & $0.113 \pm 0.101^{**}$ & $0.330 \pm 0.150^{**}$ & $0.497 \pm 0.278$ & $51.0\%$ \\
\bottomrule
\end{tabular}
\end{table*}

In terms of computational cost, PanoDiff-SR is thus five to seven times cheaper to train than the two
StyleGANs, and is also more computationally efficient than PanoDiff-HR and ADM. Of the
direct high-resolution models, only LDM and FastGAN-HR, both far behind on FID, are cheaper to train
(Table~\ref{tab:directhr}).

In summary, the comparisons show that the two-stage pipeline outperforms the direct high-resolution
diffusion models and is broadly on par with the two StyleGANs: these are ahead on FID and precision and
are harder for the classifier to detect, whereas PanoDiff-SR is ahead on recall and training cost. The
observer study was not repeated for these models, so a direct comparison by clinical readers is not
available at this point, and the value of either approach for downstream tasks remains to be determined.

\subsubsection{Synthetic radiographs versus individual source devices}
\label{sec:controls}
The training corpus used for PanoDiff-SR pools five source datasets acquired on different devices, so the synthetic
set was also compared with each device separately. Each FID was calculated five times, on random subsets of 500 images per set. FID between synthetic and real data ranged between $71.5$ (TUFTS) and $102.0$ (TSXK), a spread of
$1.43\times$, whereas FID between the real datasets ranged from $61.7$ to $117.0$, a spread of
$1.90\times$. The generator therefore does not strongly favour one device, and the device it matches least, TSXK, is also the one most distinct from the
other four (mean FID $97.1$, against $74.2$--$95.2$ for the others). Five devices are too few to
establish this relation statistically (Pearson $r=0.76$, $p=0.13$), so the result shows no
strong device bias.

\subsection{Future work}
\label{sec:future}
Several steps can build on this study. First, the U-Net size, the loss weights of the SR
stage, the number of diffusion timesteps and the mix of source datasets were taken from the
source implementations. A systematic search over such design choices has improved diffusion
models for natural images \cite{karras2022elucidating}, and it may improve this pipeline as
well. Second, the five source devices occupy different regions of the
feature space (Figure~\ref{fig:tsnefolders}). Harmonising the devices, or conditioning the
generator on the device, could reduce this source of variation \cite{guan2022domain}. Third,
each of our six observers read every image once. A larger reader panel with repeated readings
would measure the variation within each reader and would test the effect of clinical
experience \cite{obuchowski2022multireader}. Fourth, possible memorisation should be assessed in more depth, as diffusion models can copy training images in whole or in part
\cite{carlini2023extracting,somepalli2023diffusion}, so a test at the level of image patches
should precede any public release of synthetic PRs. Finally, the generator can be conditioned
on lesions such as caries or fractures, or use an anatomical mask as input, facilitating the use of synthetic data for downstream training of classification or segmentation models.

\section{Conclusion}
In this study, we developed PanoDiff-SR, an end-to-end pipeline that couples a diffusion generator (PanoDiff) with a fine-tuned transformer-based SR stage, and configured and evaluated it for the synthesis of complete dental panoramic radiographs. Apart from technical evaluation methods, we also performed a human observer study involving dentists to distinguish real from synthetic PRs, adding further support to our findings. The results indicate that the pipeline produces PRs that are difficult to identify as synthetic under time-limited clinical viewing. Because the iterative sampler operates only at low resolution, the complete pipeline trains and samples on a single GPU (Table~\ref{tab:efficiency}), which keeps full-arch synthesis at diagnostic resolution within reach of groups without large-scale computing resources.

\section{Code and Data Availability}
The implementation of both stages of PanoDiff-SR, together with the training
configurations and the pre-trained checkpoints, is available at
\url{https://github.com/s4nyam/PanoDiff}. An interactive project page at
\url{https://s4nyam.github.io/panodiff/} presents the generated radiographs at both
resolutions, the figures and tables of this paper in browsable form, and a version of the observer
task of Section~\ref{sec:expert} that the reader can attempt under the same time limit.
The five source datasets of real PRs are the public collections listed in
Table~\ref{tab:datasetoverview} and remain subject to the access conditions of their
respective providers.

\section*{Acknowledgment}
This work was funded by the Independent Research Fund Denmark, project “Synthetic Dental Radiography using Generative Artificial Intelligence”, grant ID 10.46540/3165-00237B. We thank Peter Bangsgaard Stoustrup, Ruza Bjelovucic, Louise Hauge Matzen, Ole Möbes, Bruna Neves de Freitas and Julian Woolley for participating as observers. Bruna N. de Freitas is supported by Aarhus Institute of Advanced Studies and Aarhus University Research Foundation. This project was made possible with GPU resources at the Department of Dentistry and Oral Health, Aarhus University, Denmark (accessed from 01-11-2024 to 30-04-2025), and with computational resources from the Machine Learning and Computational Intelligence (MaLeCI) Group at Aarhus University, Denmark. The authors further wish to acknowledge CSC -- IT Center for Science, Finland, and the LUMI consortium for computational resources on the LUMI supercomputer.

\bibliographystyle{IEEEtran} %
\bibliography{refs}

\end{document}